\documentclass[prb,amsmath,amssymb,lengthcheck,showpacs,floatfix,groupedaddress,superscriptaddress,longbibliography]{revtex4-1}
\usepackage{graphicx}
\usepackage[english]{babel}
\usepackage{times}
\usepackage{dcolumn}
\usepackage[usenames,dvipsnames]{color}
\usepackage{units}
\usepackage{bm}
\usepackage{soul}
\usepackage[utf8]{inputenc}
\usepackage{float}
\usepackage{dsfont} 
\usepackage{xcolor}

\begin{document}

\title{Strong-coupling theory of quantum dot Josephson junctions: role of the residual quasiparticle}

\author{Luka Pave\v{s}i\'{c}}
\email{luka.pavesic@ijs.si}
\affiliation{Jo\v{z}ef Stefan Institute, Jamova 39, SI-1000 Ljubljana, Slovenia}
\affiliation{Faculty of Mathematics and Physics, University of Ljubljana, Jadranska 19, SI-1000 Ljubljana, Slovenia}

\author{Ramón Aguado}
\email{ramon.aguado@csic.es}
\affiliation{Instituto de Ciencia de Materiales de Madrid (ICMM),
Consejo Superior de Investigaciones Cientificas (CSIC), Sor Juana Ines de la Cruz 3, 28049 Madrid, Spain}

\author{Rok \v{Z}itko}
\email{rok.zitko@ijs.si}
\affiliation{Jo\v{z}ef Stefan Institute, Jamova 39, SI-1000 Ljubljana, Slovenia}
\affiliation{Faculty of Mathematics and Physics, University of Ljubljana, Jadranska 19, SI-1000 Ljubljana, Slovenia}

\begin{abstract}
We consider an interacting quantum dot strongly coupled to two superconducting leads in a Josephson junction geometry. By defining symmetry-adapted superpositions of states from the leads, we formulate an effective Hamiltonian for the strong-hybridisation regime with a single orbital directly coupled to the dot and three additional indirectly coupled orbitals. This minimal basis set allows to account for the quasiparticles in the vicinity of the dot as well as those further away in the leads, and to describe how their role evolves as a function of coupling strength and phase bias $\phi$. This formulation also reveals the changing nature of the spin-doublet state for the experimentally relevant coupling strengths. The binding of a nearly decoupled quasiparticle in the vicinity of the QD explains the "doublet chimney" in the phase diagram for $\phi \sim \pi$, in contrast to $\phi \sim 0$ where the residual quasiparticle escapes to infinity and plays no active role.
\end{abstract}

\maketitle

\newcommand{\mycomment}[1]{}

\newcommand{\REF}[1]{\textcolor{red}{REF(#1)}}
\newcommand{\red}[1]{\textcolor{red}{#1}}
\newcommand{\rz}[1]{\textcolor{teal}{#1}}
\newcommand{\luka}[1]{\textcolor{red}{#1}}

\newcommand{\ket}[1]{\vert #1 \rangle}
\newcommand{\OS}{\ket{\mathrm{OS}}}
\newcommand{\MO}{\ket{\mathrm{MO}}}

\newcommand{\localS}[2]{\mathcal{S}_{#1}^{#2}}

\newcommand{\bondingS}{S_b}
\newcommand{\bondingD}{D_b}
\newcommand{\id}{\mathds{1}}

\section{Introduction}

Josephson junctions (JJs) are key constituents in modern platforms for quantum state engineering using superconductors (SCs) \cite{Devoret-Schoelkopf} and they are continuously enhanced with novel functionalities such as gate tunability, compatibility with magnetic fields \cite{Aguado:APL20}, and unconventional Josephson potentials for building parity-protected qubits \cite{Larsen:2020}. Another target is fully exploiting the microscopic degrees of freedom in the JJ, namely its subgap levels in the few-channel regime. A recent example are Andreev spin qubits (ASQ) \cite{Zazunov2003,Chtchelkatchev:2003,Padurari:2010}, where quantum information is stored in the spin degree of freedom of a trapped quasiparticle. The first experimental realisation of this idea \cite{Hays2021} relied on non-interacting Andreev levels, but required a complex scheme involving higher energy levels for qubit manipulation. A promising alternative makes use of interacting subgap states in JJs with an embedded semiconducting quantum dot (QD) tuned into a spinful doublet ground state (GS) \cite{Bargerbos2022,Bargerbos:22b,Pita-Vidal:2022}. The QD needs sufficiently strong coupling to the SC leads for efficient control and read-out, but at too large coupling the doublet is no longer the GS, leading to increased leakage out of the computational subspace. The competition between singlet and doublet GSs is governed by the coupling strength, charging energy, QD filling, and phase difference across the junction, $\phi$ \cite{Kirsanskas2015,Meden2019}. Furthermore, a doublet GS that is appropriate for use in ASQs needs to fulfill specific requirements, in particular good manipulability using local electromagnetic fields produced by nearby gate electrodes \cite{Pita-Vidal:2022}. For this reason, not only the phase diagram is important, but also the doublet state wavefunction and its properties.

\begin{figure}
    \centering
    \includegraphics[width=0.9 \columnwidth]{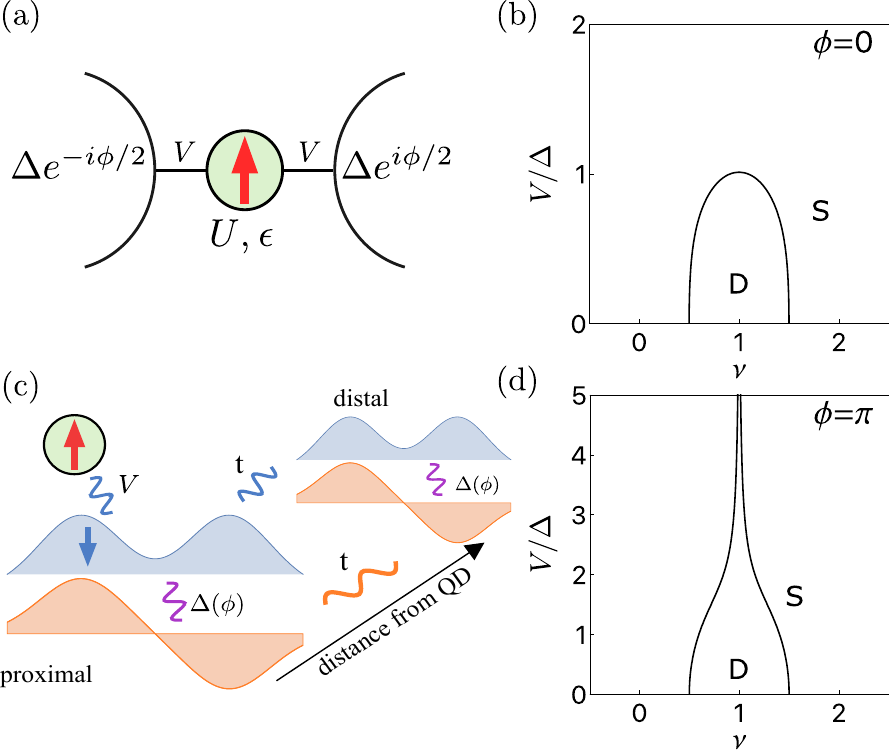}
    \caption{
    (a) Sketch of the quantum dot Josephson junction. 
    (b, d) Phase diagrams for $\phi=0$ and $\phi=\pi$ within the minimal model, Eqs.~\eqref{eq:hop} and \eqref{eq:SC}, for a left-right symmetric system. $V$ is the coupling strength, $\nu = 1/2 - \epsilon/U$ is the dot filling in units of charge. Parameters are $U=10\Delta$, $t=0.2\Delta$.   
    The ``doublet chimney'' for $\phi=\pi$ is given in the ZBA limit by $|\nu-1|=\Delta^2/4V^2$, Eq.~\eqref{chimney}.
    (c) Sketch of the model expressed in terms of a proximal and a distal set of symmetric and antisymmetric superconductor orbitals.
    Only the proximal symmetric orbital couples directly to the dot.
    Superconducting pairing materializes as a $\phi$-dependent anomalous hopping that mixes the different types of orbitals, as described in the Hamiltonian in Eq.~\eqref{eq:SC}. The hopping term $t$ simulates the finite bandwidth of the superconductors.
    }
    \label{fig:1}
\end{figure}

Optimized ASQs will most likely operate in the intermediate coupling regime. This regime is easily accessible using modern impurity solvers and it has been rather thoroughly explored using the numerical renormalization group (NRG) and other methods. For the physical interpretation of the obtained numerical results one usually relies on simplified toy models defined on smaller Hilbert spaces. Our work introduces an effective model with a minimal set of orbitals that provides qualitatively correct results for all coupling strengths, from weak to strong coupling regimes, and for all values of $\phi$. Furthermore, our work clarifies the role of Bogoliubov quasiparticles in the superconducting lead in the formation of the subgap states. In particular, we point out the need for taking into consideration not only the quasiparticles in the immediate vicinity of the quantum dot, but also those further away in the leads (but not necessarily in infinity, where they may be ignored altogether). This work hence goes beyond the approximations such as the superconducting atomic limit (SAL) and the zero-bandwidth approximation (ZBA) \cite{affleck2000, Vecino2003, bergeret2007,GroveRasmussen2018, twoBogoliubov,Zonda2022, Hermansen2022, Schmid2022}, defined only on local orbitals (QD orbital alone for SAL, QD orbital + one orbital per lead in ZBA). It does so by including the leading order effects of the finite bandwidth in the superconducting channels. This step is necessary to correctly capture all properties of the subgap states for strong coupling, in particular their symmetry properties.

We approach this task by reformulating the superconducting Anderson model (SAM) in the basis of symmetry-adapted orbitals, using a minimal set of states that permits to account for the finite bandwidth. The model makes it possible to properly account for the screening of the QD spin, the main mechanism responsible for lowering the energy of the subgap states below the continuum of elementary quasiparticle (Bogoliubov) excitations. It has long been believed that the QD spin is fully screened in the singlet and ``unscreened'' in the doublet GS, but NRG results show that the doublet state is actually partially screened for any non-zero value of coupling \cite{Moca2021,knight}. In fact, for very strong coupling the QD spin is completely screened in all states, singlet and doublet, for all values of $\phi$ \cite{knight}. 

In this work, we show that in the strong-coupling limit the doublet becomes the same as the well-understood singlet state, but with an overall doublet character owing to one residual spin resulting from the broken Cooper pair. We furthermore establish that the wavefunction of this spin (its orbital character) depends on the value of $\phi$: for $\phi\sim 0$ it is far away from the QD in a symmetric orbital, while for $\phi \sim \pi$ it is located closer to the QD in an antisymmetric orbital. Importantly, while the free magnetic moment exists in the doublet state at large coupling, it is no longer localized on the QD itself, but rather smeared across the superconducting leads. The exact spatial location and extent depend on the model parameters, especially the phase bias $\phi$. This puts constraints on the coupling strength in JJs intended to be used as ASQs: if the coupling is too strong, the moment is less responsive to modulation and readout schemes that locally address the dot. Furthermore, the qubit encoded in the spin degree of freedom of the ASQ is expected to have different decoherence rate depending on the distribution in space of the spin doublet wavefunction because the QD and the superconducting leads are made of materials representing different noise environments.

The notion of unscreened QD spin in the doublet states is also challenged by the existence of the ``doublet chimney''.
This extended doublet phase at $\phi=\pi$ that persists even for large coupling strengths where a singlet GS is generally expected [see Fig.~\ref{fig:1}(d)] has been predicted theoretically \cite{Shiba1969, Matveev1993, Rozhkov1999, Clerk2000, Siano2004} long ago and recently directly observed in experiment \cite{Bargerbos2022}. We show that this limit (exactly  at $\phi=\pi$) is actually correctly captured by the ZBA. Expansion in the inverse coupling strength gives analytical insight into the regime of intermediate QD-SC coupling and explains the role of the residual spin as well as the shape of the phase boundary.

\section{Model}
\label{sec:model}

The system, sketched in Fig.~\ref{fig:1}(a), is modelled by SAM, $H = H_\mathrm{SC}^{(L)} + H_\mathrm{SC}^{(R)} + H_\mathrm{QD} + H_\mathrm{hop}$, with
\newcommand{\Vbeta}{V_\beta}
\newcommand{\VL}{V_L}
\newcommand{\VR}{V_R}
\begin{equation*}
\begin{split}
    H_\mathrm{SC}^{(\beta)} &= \sum_{k\sigma} \epsilon_k c^\dagger_{\beta k\sigma}c_{\beta k \sigma} - \Delta \sum_k e^{i \phi_\beta} c^\dagger_{\beta k \downarrow}c^\dagger_{\beta k\uparrow} + \text{H.c.}, \\
    H_\mathrm{QD} &= \epsilon \sum_\sigma \hat{n}_{d\sigma} + U \hat{n}_{d\uparrow}\hat{n}_{d\downarrow}, \\
    H_\mathrm{hop} &= -\frac{1}{\sqrt{\mathcal{N}}} \sum_{\beta = L,R} \Vbeta \sum_{k\sigma} d^\dagger_\sigma c_{\beta k\sigma} + \mathrm{H.c.} \\
\end{split}
\end{equation*}
$c_{\beta k\sigma}$ is the operator for an electron in the superconductor $\beta \in \{ L,R \}$ with energy $\epsilon_k$ and spin $\sigma$. $d_\sigma$ is the operator for the QD level, $\hat{n}_{d\sigma} = d^\dagger_\sigma d_\sigma$ the corresponding number operator, $\epsilon$ the impurity level, $U$ the on-site interaction, and $V_\beta$ the hopping to lead $\beta$. 
We parameterize the hoppings as $\VL = V(1-\eta)$, $\VR=V(1+\eta)$, so that $\eta$ quantifies the left-right asymmetry.
$\mathcal{N}$ is the number of $k$-states in a SC lead. We set $\phi_L=-\phi/2$ and $\phi_R=\phi/2$.

We simplify the Hamiltonian in several steps. In the first step, we apply the gauge transformation $c_{\beta k \sigma} \to e^{\phi_\beta/2} c_{\beta k\sigma}$, \cite{Oguri2004,Choi2004}.  This removes the phase from the pairing terms and transfers it to the hybridisation part, $H_\mathrm{hop}$.

In the second step, we reduce the infinite basis set by retaining two states for each superconductor, one representing states in the immediate vicinity of the QD, another representing states far away from the QD. One could, in principle, determine the set of the most relevant orbitals numerically, for example by determining the natural orbitals of the impurity problem (we return to this question in the conclusion). For simplicity, we here choose instead two localized orbitals, $c_{\beta\sigma}(r)=(1/\sqrt{\mathcal{N}})\sum_k e^{ikr} c_{\beta k\sigma}$ with $r=0$ and $r=l$, respectively. By transforming the kinetic-energy part of the Hamiltonian to the new basis, we find that the two orbitals are coupled by a complex-valued hopping term obtained by Fourier transforming the dispersion, $t = (1/\mathcal{N}) \sum_k e^{ikl} \epsilon_k$.
The parameter $l$ has no particular physical meaning, and only the resulting $t$ has a bearing on the effective Hamiltonian. For an orbital far from the QD (large $l$, in particular much larger then the Fermi wavelength $2\pi/k_F$) the sum rapidly oscillates and $t$ is small. We will consider $t$ to be a free parameter that is much smaller than the bare bandwidth, but non-zero. The qualitative behavior of the results does not depend on the value of $t$ (see Appendix~\ref{appA}), what matters above all is that $t$ accounts for the non-zero mobility of quasiparticles (QPs) in the truncated model.

In the final step, we define the orthogonal symmetric $b$ and antisymmetric $a$ orbitals:
\begin{equation*}
\label{eqba}
\begin{split}
    b_{\sigma}(r) &= \frac{1}{\sqrt{\VL^2+\VR^2}} \left[ \VL e^{-i\frac{\phi}{4}} c_{L\sigma}(r) + \VR e^{i\frac{\phi}{4}} c_{R\sigma}(r) \right],\\
    a_{\sigma}(r) &= \frac{1}{\sqrt{\VL^2+\VR^2}} \left[ -\VR e^{-i\frac{\phi}{4}} c_{L\sigma}(r) + \VL e^{i\frac{\phi}{4}} c_{R\sigma}(r)\right]
\end{split}
\end{equation*}
resulting in a minimal model $H=H_\mathrm{SC}+H_\mathrm{QD}+H_\mathrm{hop}$ where
\begin{equation}
    H_\mathrm{hop} =  - V \sqrt{2 \left( 1+\eta^2 \right)} \sum_\sigma \left[ d^\dagger_\sigma b_{\sigma}(0) + \mathrm{H. c.} \right]
\label{eq:hop}
\end{equation}
describes the coupling between the QD and the symmetric
proximal orbital $b(0)$, while
\begin{equation}
\begin{split}
    H_\mathrm{SC} =& -\Delta \cos\frac{\phi}{2} \sum_{r=0,l} \left[ a_{\downarrow}(r) a_{\uparrow}(r) + b_{\downarrow}(r) b_{\uparrow}(r) + \mathrm{H. c.} \right] \\
     -& i \Delta \frac{1 - \eta^2}{1 + \eta^2} \sin\frac{\phi}{2} \sum_{r=0,l} \left[ a_{\uparrow}(r) b_{\downarrow}(r) - a_{\downarrow}(r) b_{\uparrow}(r) + \text{H.c.} \right] \\
     -& i \Delta \frac{2\eta}{1+\eta^2} \sin\frac{\phi}{2} \sum_{r=0,l} \left[ b_{\downarrow}(r) b_{\uparrow}(r) - a_{\downarrow}(r) a_{\uparrow}(r)  + \mathrm{H.c.} \right] \\
     -& t \sum_\sigma \left[ a^\dagger_{\sigma}(0) a_{\sigma}(l) + b^\dagger_{\sigma}(0) b_{\sigma}(l) + \mathrm{H.c.} \right]
\label{eq:SC}
\end{split}
\end{equation}
describes the additional effects brought about by possible additional quasiparticles in the system.

The special cases where the model further simplifies are:
\begin{itemize}
    \item $\phi=0$, where the second and third lines are zero;

    \item $\phi=\pi$, where the first line is zero;

    \item $\eta=0$, where the third line is zero;

    \item $\phi=\pi$ and $\eta=0$, i.e., the combination of the former two.
\end{itemize}
In the first case, the model has time-reversal invariance, with $d_\uparrow \to -d_\downarrow$, $d_\downarrow \to d_\uparrow$, followed by complex conjugation, which inverts spin.
In the second case,  the system has a different symmetry: $d_\uparrow \to d_\downarrow$, $d_\downarrow \to d_\uparrow$ (without any sign change in the operators) followed by complex conjugation. This operation reflects spin across the $(xy)$ plane in the spin space. It is an antiunitary symmetry different from the time-reversal symmetry at $\phi=0$.
The third case is by definition the left-right (LR) symmetric situation.
Finally, the fourth case corresponds to a particularly high symmetry: this is the regime where the doublet chimney persists to arbitrarily large coupling strength and the singlet is never the GS.

We note that for $t=0$, our model becomes equivalent to the zero-bandwidth approximation (ZBA) amended with additional orbitals that are fully decoupled from the QD (i.e., $l\to\infty$ limit). Alternatively, our model can be seen as the minimal extension of the ZBA taking into account the finite bandwidth of the superconductors. Comparing the results of our model for finite $t$ and for strictly zero $t$ (i.e., ZBA amended with additional fully decoupled orbitals), we find similar results except for the lifting of level degeneracies at non-zero $t$, and for the discontinuous evolution (vs. V) at zero $t$ instead of the smooth cross-overs at non-zero $t$; the solution at non-zero $t$ has the same qualitative behavior as the NRG solution of the full Hamiltonian, motivating the choice of finite value of $t$. (In principle, one could also fix $t$ by comparing with the reference NRG results.)
See Sec.~\ref{ZBA} for further discussion.

\section{Eigenstates}

The strong-coupling theory is based on expanding in $1/V$, with $H_\mathrm{hop}$ as the non-perturbed part, and $H'=H_\mathrm{QD} + H_\mathrm{SC}$ as the perturbation. We use projector-based perturbation theory (PT) to deal with the large degeneracy \cite{Yao2000}. 

Low-energy eigenstates of $H_\mathrm{hop}$ have two electrons occupying the $d-b(0)$ bonding orbital,
\newcommand{\CB}{|\mathrm{B}\rangle}
\begin{equation}
\label{CB}
    \CB = \frac{d^\dagger_\uparrow + b^\dagger_{\uparrow}(0)}{\sqrt{2}} \, \frac{d^\dagger_\downarrow + b^\dagger_{\downarrow}(0)}{\sqrt{2}} \ket{0},
\end{equation}
with a bonding energy of $E_B = - 2 V \sqrt{2\left(1+\eta^2\right)}$. The configuration of electrons in the remaining SC orbitals is arbitrary and does not affect the energy of unperturbed states. It does determine their total spin, with the doublets having a free spin in one of the orbitals.

For non-zero $1/V$ the GS degeneracy is lifted. The first-order energy corrections are
\begin{equation}
\begin{split}
    \label{eq:first_order} 
    \delta E^{(1)}_\mathrm{S} &= \frac{U}{4} + \epsilon \pm \Delta \frac{\sqrt{ \left[\left(1+\eta^2\right)\cos\frac{\phi}{2} \right]^2 + \left[ 2\eta \sin \frac{\phi}{2} \right]^2}}{1+\eta^2} ,\\
    \delta E^{(1)}_\mathrm{D} &= \frac{U}{4} + \epsilon,
\end{split}
\end{equation}
for the lowest singlet (S) and the lowest doublet (D). The $\Delta$-term simplifies to $ \delta E^{(1)}_\mathrm{S} = \frac{U}{4} + \epsilon \pm \Delta\cos(\phi/2)$ for $\eta=0$, the cosine factors originating from the interference processes between the two leads; this is the same cosine factor that arises in the anomalous part of the hybridisation function of SAM \cite{kadlecova2017}.
Unless $\eta=0$ and $\phi=\pi$, the two S states are split, with D exactly midway between them.
Finite bandwidth effects ($t$ term) favor states where the $a$ orbitals form an $a(0)-a(l)$ bond.
In the doublet state this leaves the $b(l)$ orbital occupied by a free spin, while in the singlet states two types of a local singlet are formed in the $b(l)$ orbital, namely $\id_{b(l)} \pm b^\dagger_{\downarrow}(l)b^\dagger_{\uparrow}(l)$.
The sum with equal phase (+ sign) for the lower-energy state can be interpreted as a Cooper pair, while the sum with the opposite phase (-) for the higher-energy state corresponds to a broken Cooper pair, i.e., two QPs.

\begin{figure}[htbp]
    \centering
    \includegraphics[width= \columnwidth]{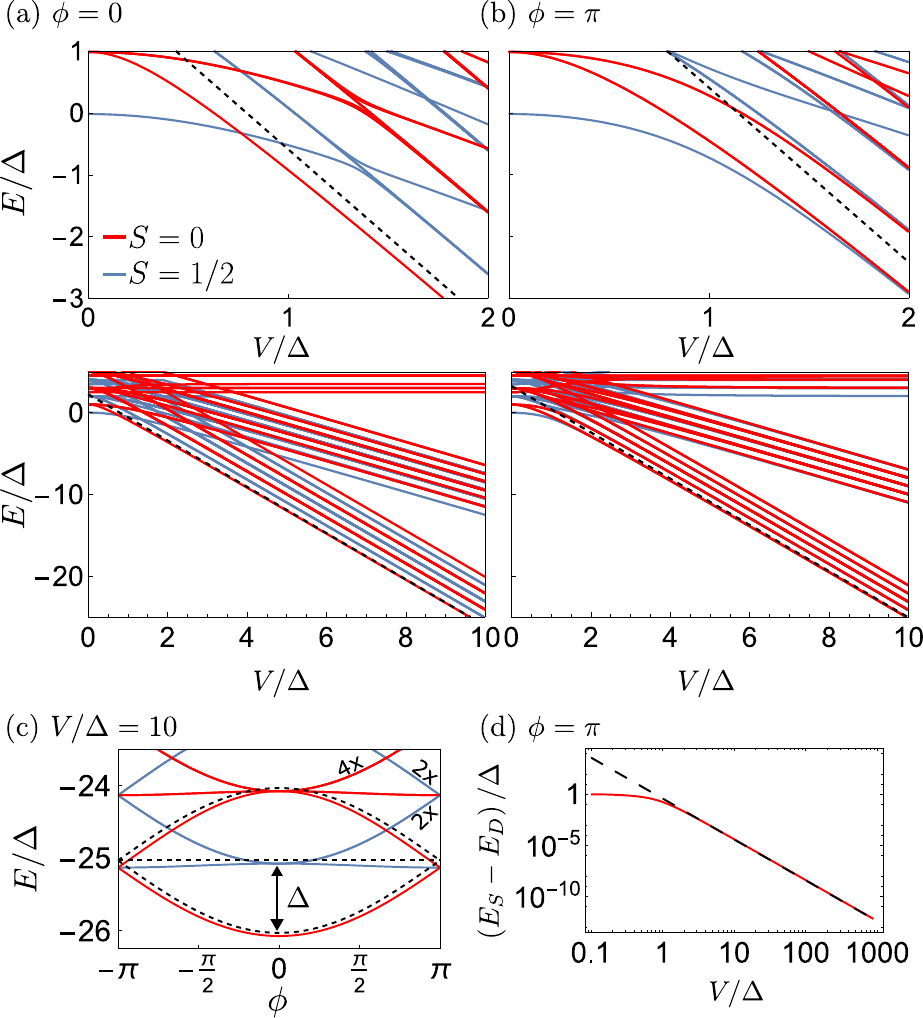}
    \caption{
    Low-energy spectra, singlet (red, $S=0$) and doublet (blue, $S=1/2$) states.
    (a) $\phi=0$ and (b) $\phi=\pi$ eigenenergies vs coupling strength $V$.
    The top row shows the spectra at small to intermediate $V/\Delta$ and the bottom row shows the spectra up to large $V/\Delta$.
    The black dashed lines are the first-order expansion for large $V$, Eq.~\eqref{eq:first_order}.
    (c) 
    $\phi$-dependence at large $V/\Delta$.
    Full lines: eigenvalues of the Hamiltonian, dashed lines: first-order expansion for large $V$, Eq.~\eqref{eq:first_order}. The labels $2\times$ and $4\times$ indicate the double or quadruple multiplicity of the excited states.
    (d) Energy difference between the lowest doublet and singlet at $\phi=\pi$ in log-log scale. Dashed line: fifth-order correction, Eq.~\eqref{eq:fifth_order}.
    Parameters are $\eta=0$, $U=5 \Delta$, $\epsilon=-U/2$, $t=0.1\Delta$. The energies are shifted by $U/2+4\Delta$ so that the zero-coupling ground state energy is at zero.
    }
    \label{fig:2}
\end{figure}

The energy spectrum is shown in Fig.~\ref{fig:2}.
At small to intermediate $V\sim\Delta$, the singlet becomes the ground state for $\phi=0$ as the bonding energy of $\CB$ overcomes the energy penalty of breaking of a Cooper pair.
In the doublet sector, we find an avoided crossing as the nature of the lowest doublet transforms from the state with a decoupled QD spin at $V\rightarrow 0$ into the state with a large contribution of $\CB$ at large $V$.
For $\phi=\pi$ the doublet remains below the singlet state for all $V$, resulting in the formation of the chimney. 
At $V \ll \Delta$ the spectrum splits into several manifolds depending on the occupancy of the bonding and antibonding orbitals between $d$ and $b(0)$, with all states within the same manifold having the same $V$-dependence.

Fig.~\ref{fig:2}(c) shows the $\phi$-dependence in the lowest manifold at large $V$. In the singlet subspace (red), the GS contains a Cooper pair, gaining $\Delta \cos(\phi/2)$ condensation energy, while the first excited singlet has one broken Cooper pair, costing an additional $\Delta \cos(\phi/2)$. The lowest-energy doublet (blue) contains a single QP. This pattern repeats with higher excitations containing increasing numbers of broken Cooper pairs and QPs \cite{combescot}.

\section{Spin screening}

At $V=0$, the GS is a doublet $D_0$, a product state composed of a free spin on the QD and decoupled SC leads. At $V=\infty$, the GS is a doublet $\bondingD$, where the QD spin is bound into a singlet with the quasiparticle in orbital $b(0)$ in the same way as in the singlet state, and a residual quasiparticle in orbital $b(l)$. In true ZBA this state cannot be represented because the Hilbert space is simply too small: the addition of an orbital such as $b(l)$ is necessary for this state to be even defined.

\begin{figure}
    \centering
    \includegraphics[width=1.0 \columnwidth]{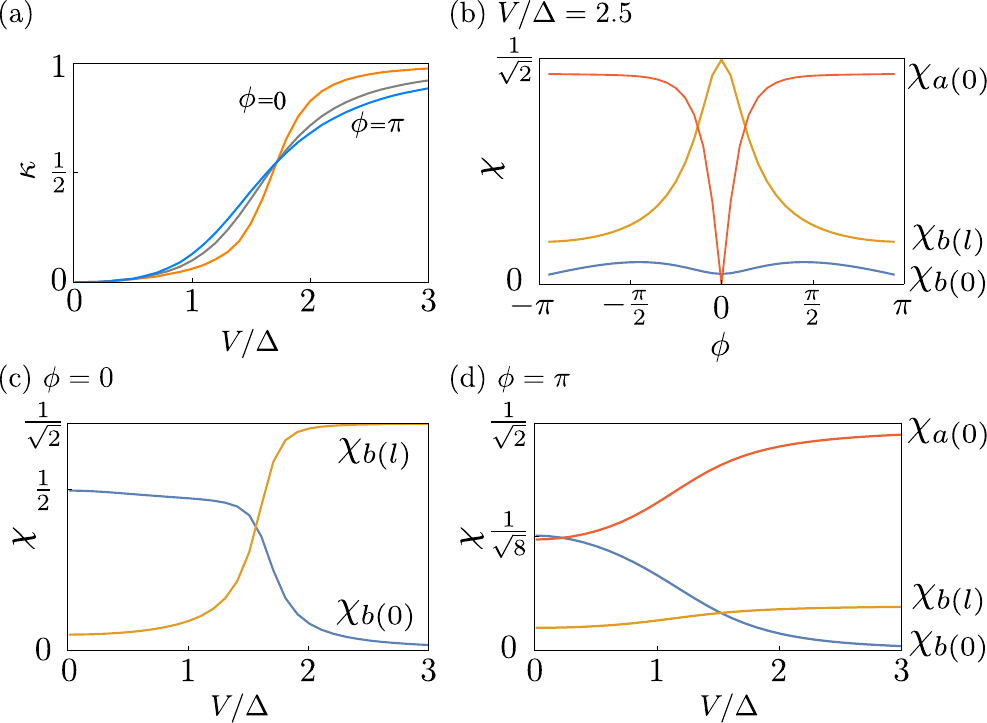}
    \caption{
    Local moment screening in the doublet ground state.
    (a) Spin compensation $\kappa$ vs $V$ for $\phi=0$ (yellow), $\pi/2$ (magenta), and $\pi$ (blue). 
    (b) Addition amplitudes $\chi$ vs $\phi$ for $V=2.5\Delta$.
    (c) Addition amplitudes $\chi$ vs $V$ for $\phi=0$.
    (d) Addition amplitudes $\chi$ vs $V$ for $\phi=\pi$. 
    Parameters are $\eta=0$, $U=10\Delta$, $\epsilon=-U/2$, $t=0.2\Delta$. $\chi_{a(l)}$ is negligibly small in all cases.
    }
    \label{fig:screening}
\end{figure}

The mixing of states $D_0$ and $D_b$ as a function of $V$ can be quantified using the spin compensation 
\begin{equation}
    \kappa = 1 - 2\langle S^z_\mathrm{QD} \rangle,
\end{equation}
ranging from $\kappa=0$ for a free spin to $\kappa=1$ for a completely screened QD \cite{Moca2021}. 
Fig.~\ref{fig:screening}(a) shows that $\kappa(V)$ is indeed monotonously increasing from 0 to 1 as $V$ is increased. It weakly depends on $\phi$ due to anomalous hopping terms in Eq.~\eqref{eq:SC}. 
We note that for zero $t$, the curves for $\phi \neq \pi$ are not smoothly increasing, but show a discrete jump where the doublet ground state and the first excitation cross (see Sec.~\ref{ZBA} for a comparison of finite and zero $t$). 
Devising a minimal model which qualitatively correctly captures the evolution of the lowest-energy doublet state from $D_0$ to $D_b$ character, for all values of $V$ and $\phi$, is the first main result of this work.

The relation between the lowest singlet and doublet states is revealed through matrix elements 
\begin{equation}
    \chi_{\alpha} = \vert \langle D \vert \alpha^\dagger_\uparrow \vert S \rangle \vert
\end{equation}
with $\alpha \in \{a(0), a(l), b(0), b(l)\}$.
Fig.~\ref{fig:screening}(b) shows the $\phi$ dependence of $\chi$ at large $V=2.5\Delta$. 
Because the QD spin is always completely screened in the singlet, $\chi_{b(0)}$ is small when the $(d,b(0))$ configuration in both states is similar,
i.e., when the QD spin is screened in the doublet as well.
The remaining $\chi$ quantify the position of the spin-carrying residual QP in the doublet. For $\phi \sim 0$, it clearly resides in the orbital $b(l)$. The maximum value $1/\sqrt{2}$ is explained by the fact that the state of $b(l)$ in the singlet is $\left( \id_{b(l)} + b^\dagger_{\downarrow}(l)b^\dagger_{\uparrow}(l)\right)/\sqrt{2}$. For $\phi \sim \pi$, the residual spin resides mostly in $a(0)$, with some weak admixture of $b(l)$; the ratio depends on $t/\Delta$, and it goes to zero as $t\to0$.
Fig.~\ref{fig:screening}(c) shows the $V$ dependence at $\phi=0$. It directly confirms the interpretation of $\kappa$ variation in terms of the changing nature of the doublet from $D_0$ to $D_b$. Fig.~\ref{fig:screening}(d) shows the $V$ dependence at $\phi=\pi$. At $V=0$, the singlet is a linear combination of singlet states involving $a(0)$ and $b(0)$, seen through equal values of the corresponding $\chi$. With increasing $V$, the singlet and doublet evolve into a similar screened state, expect for the residual quasiparticle in orbital $a(0)$.

We observe that $\chi_a(l)$ is always negligibly small. This simply indicates that this orbital plays no role
in spin screening. Nevertheless, it needs to be retained in the Hamiltonian in order to obtain the correct energetics (i.e., the order of states) by the formation of a $a(0)-a(l)$ valence bond states, as discussed in the previous section.

The ZBA is commonly used to explore the subgap spectrum, and it is widely believed to reproduce all qualitative features correctly. For $\phi=0$, ZBA  predicts a level crossing between two doublet states having different mirror symmetry  as $V$ is increased, which is at odds with the NRG calculations for the full model \cite{rok_zitko_2022_5874832} that clearly show that at $\phi=0$ the symmetry of the doublet state (which is a mixture of $D_0$ and $D_b$) does not change.
In ZBA with no distal orbitals, the second QP of the doublet state at high $V$ is constrained to always sit in the proximal antisymmetric orbital $a(0)$, while in our model the finite hopping $t$ allows it to tunnel to a distal symmetric orbital $b(l)$, while $a(0)$ and $a(l)$ form an inter-site singlet, thereby reducing the energy of $D_b$ and restoring the expected symmetry as well as the correct order of states at $\phi=0$. This explains the origin of the main deficiency of ZBA in the strong-coupling limit for values of $\phi$ close to 0.

The identification of the orbital in which the residual quasiparticle resides is the second main result of this work. Our model is thus indeed the minimal model that can qualitatively describe the fate of both quasiparticles following the breaking of the Cooper pair by the magnetic impurity.

\section{Doublet chimney}

At $\phi = \pi$ and $\eta=0$, the lowest-lying singlet and doublet states are degenerate to lowest order, see Eq.~\eqref{eq:first_order}. 
Yet, the exact numerical solution gives slightly lower energy for the doublet. 

It is possible to analytically calculate high-order corrections at $\phi=\pi$ because the ``nearly free'' quasiparticle occupies a proximal orbital $a(0)$, see Fig.~\ref{fig:screening}(b,d), thus it is admissible to set $t=0$ without qualitatively changing the low-energy states. For $\eta=0$ the lowest two singlet states are exactly degenerate at $\phi=\pi$ as $H'=H_\mathrm{QD}+H_\mathrm{SC}$ does not mix them in any order. This is the result of the fourth symmetry case discussed in Sec.~\ref{sec:model}. In the following we focus on this special case of $\phi=\pi$ and $\eta=0$, where the ZBA is an adequate description.

We now use the non-degenerate Rayleigh-Schrödinger PT. In third order, we find 
\begin{equation}
    \delta E^{(3)}_{S} - \delta E^{(3)}_{D} = -\frac{1}{2}\frac{U \Delta^2}{V^2} \left\vert \nu - 1 \right\vert,
    \label{eq:tri}
 \end{equation}
with $\nu = \frac{1}{2} - \frac{\epsilon}{U}$ the QD filling in units of particle number.
For $\nu=1$, a difference is found only in the fifth order and it has a surprisingly simple form:
\begin{equation}
    \delta E^{(5)}_{S} - \delta E^{(5)}_{D} = \frac{1}{8} \frac{U\Delta^4}{V^4}.
    \label{eq:fifth_order}
\end{equation}
The energy corrections at all lower orders are exactly the same in both spin sectors. At fifth order, there are $U^5$, $U^3\Delta^2$ and $U\Delta^4$ contributions,
with only the terms of the last kind not cancelling out. Fig.~\ref{fig:2}(d) shows that Eq.~\eqref{eq:fifth_order} becomes a good approximation for $V\gtrsim\Delta$. 
Combining the third and fifth order equations gives the shape of the transition line (the chimney) as 
\begin{equation}
\label{chimney}
|\nu-1|=\Delta^2/4V^2.
\end{equation}

At large $V$, the $d$ and $b(0)$ orbitals are equally strongly coupled in S and D states and, counter-intuitively, it is the state of the antisymmetric orbital $a(0)$ (the number of quasiparticles it contains) that differentiates them via higher order processes. This residual interaction is a kind of blocking effect \cite{vonDelft2001}: pairing processes are ineffective for orbitals occupied by a single quasiparticle,
leading to different combinatorial prefactors that result in the non-zero fifth order energy difference.
\footnote{All perturbation calculations as well as an extended collection of exact numerical results is available in the form of Wolfram Mathematica notebooks in the Supplemental material.}
Obtaining a deeper understanding of the physical origin of the doublet chimney is the third main result of this work.

The energy difference only appears for $U \neq 0$, see Eq.~\eqref{eq:fifth_order}.
This confirms that the doublet is stabilized by an effective interaction between the QP screening the QD local moment and the free QP.
In the non-interacting resonant limit ($U=0$, $\epsilon=0$) the completely proximitized QD level is occupied by a Cooper pair (Andreev bound state) and there is only one free QP in the doublet state. At $\phi=\pi$ the singlet and doublet states are then exactly degenerate.
An experimental observation of a doublet chimney thus directly implies an interacting QD level.

We note that an analogous phenomenon of persistent doublet GSs is also found in the hard-gap Anderson impurity model \cite{takegahara1992,CJ,galpin2008,galpin2008epj,Moca2010,Zalom2021,Zalom2022}, to which the QD JJ problem maps for $\phi=\pi$ \cite{Zalom2021,Zalom2022}. Such states have been interpreted in terms of a fixed-point effective Hamiltonian obtained in an NRG analysis of finite-size spectra \cite{CJ} and through the analytical structure of the impurity self-energy which features a $\delta$-peak pinned at the Fermi level \cite{galpin2008,galpin2008epj}.
Interestingly, the fixed-point analysis in Ref.~\onlinecite{CJ} was also based on a two-orbital description \footnote{Eq.~(8) in Ref.~\onlinecite{CJ}, involving operators $f$ and $g$.}. 

If the mirror symmetry is broken ($\eta\neq0$), the doublet chimney no longer extends to infinite hybridisation strength and it disappears altogether for large asymmetry (see also App.~\ref{appB}).

\section{Discussion: relation to the zero-bandwidth approximation}
\label{ZBA}

\begin{figure}[htbp]
    \centering
    \includegraphics[width= \columnwidth]{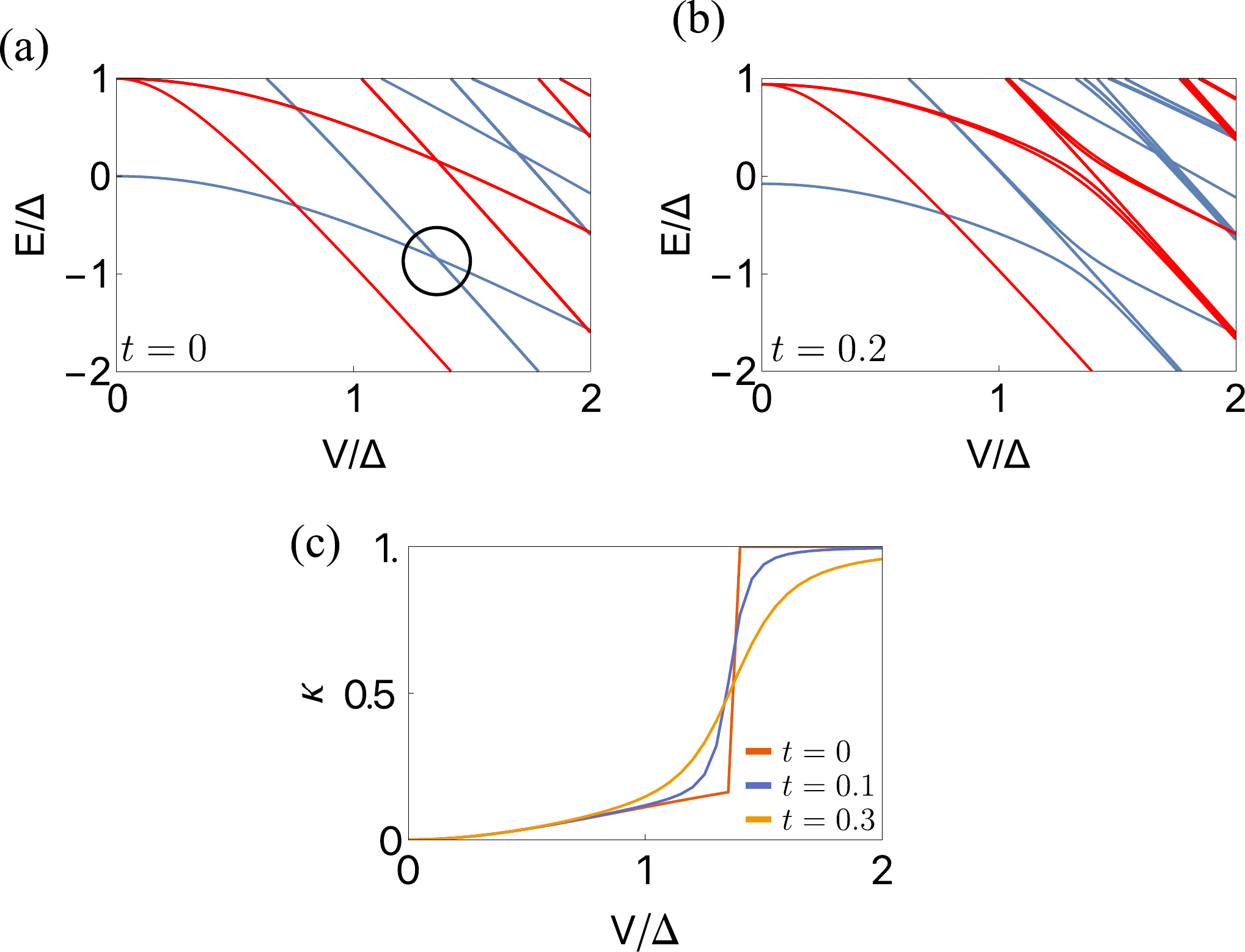}
    \caption{
    The effect of $t$ on the doublet states at $\phi = 0$.
    (a) Zoom-in on the $V$ dependence of the spectra shown in Fig.~\ref{fig:2}(a) for $t=0$. The black circle shows the crossing of the first excited ($D_b$, degenerate with $D_a$) and the ground ($D_0$) doublet states.
    (b) Same plot for $t=0.1$. The mixing between the states $D_b$ and $D_0$ pushes one linear superposition of these mirror-symmetric doublets below the mirror-antisymmetric $D_a$.
    (c) Spin screening $\kappa$ for the same parameters and different $t$.
    }
    \label{fig:t_spectra_zoom}
\end{figure}

In Fig.~\ref{fig:t_spectra_zoom}, we show the striking difference in spin screening $\kappa$ for $t=0$ and finite $t$. 
The spectra in Fig.~\ref{fig:t_spectra_zoom}(a) at $t=0$ shows the crossing of the lowest doublet ($D_0$) and the first two excited doublet states ($D_b$, $D_a$) at $V/\Delta \sim 1.3$ (black circle). Here, $D_b$ is the screened doublet introduced before, while $D_a$ is a mirror-asymmetric (ungerade) screened doublet state, with one quasiparticle in the $b(0)$ orbital and another quasiparticle in the $a(0)$ orbital. In true ZBA, $D_b$ state does not exist, there is only $D_a$, because $b(l)$ does not exist.

For finite $t$, the crossing between $D_0$ and $D_b$ becomes an avoided crossing, with the two states mixing, see Fig.~\ref{fig:t_spectra_zoom}(b). The asymmetric orbital $D_a$ is unaffected. This results in the lowest doublet state always having the correct symmetry (mirror symmetric, gerade). Furthermore, the degree of screening in the lowest doublet is then continuously increasing, consistent with the NRG results.
This is reflected in $\kappa(V)$ in the doublet ground state, Fig.~\ref{fig:t_spectra_zoom}(c).
For $t=0$, it exhibits a discrete jump at the point of the doublet state crossing, as the completely screened $D_b$ (or the degenerate $D_a$) becomes the ground state. 
The mixing, induced by $t$, smoothens the jump into the expected crossover.

\section{Conclusion}

Except for certain high-symmetry points, the partially screened doublet states in the strong-coupling limit of quantum impurity Hamiltonians with a gapped bath cannot be reproduced with a single-orbital (ZBA) description of the bands. Our work explains this requirement in the context of superconducting systems from the perspective of the two electrons following the break-up of the Cooper pair due to exchange coupling to the QD: one electron forms the singlet state with the QD local moment, while the other either experiences residual interaction (to produce a doublet ground state at $\phi \sim \pi$) or goes away to large distances (to produce the doublet excited state at $\phi \sim 0$) or infinity (to produce the singlet ground state at $\phi \sim 0$). Single-orbital descriptions cannot cover all these possibilities.

The multi-orbital approximation proposed in this work provides insights into the spin-screening mechanisms and the nature of the doublet state for all coupling strengths and all values of phase bias $\phi$. That will be instrumental in the design of complex hybrid devices based on coupled spins and SC degrees of freedom, such as ASQs. 
For example, our results imply that the doublet spin can be partially redistributed into the SCs, and thus cannot be manipulated by experimental methods locally addressing the QD.
We also explained the curious doublet chimney as a kind of blocking effect arising from the presence of a residual quasiparticle, located close to the QD at $\phi = \pi$.

A possible follow-up to this work would be a detailed real-space study of the relevant quasiparticle wave-functions. A possible path would be through constructing the natural orbitals \cite{Lowdin1955,Lu2014,He2014,Bi2019,debertolis2021} after appropriate generalization for the superconducting case \cite{Schmidt2019}. Experimentally, the localization properties of the doublet state wavefunction are revealed, for example, through the impurity Knight shift effect --- the effective $g$-factor of the QD level is $\phi$-dependent \cite{knight}.
It would be of interest to perform such measurements in a controlled manner, i.e., as a function of coupling strength $V$ and phase bias $\phi$.

\begin{acknowledgments}
LP thanks G\'erman Blesio and Szczepan Głodzik for discussions and ideas.
We acknowledge the support of the Slovenian Research and Innovation Agency (ARIS) under P1-0416 and J1-3008 and of the Spanish Ministry of Science through Grants PID2021- 125343NB-I00 and TED2021-130292B-C43 funded by MCIN/AEI/10.13039/501100011033, "ERDF A way of making Europe" and European Union NextGenerationEU/PRTR. Support by the CSIC Interdisciplinary Thematic Platform (PTI+) on Quantum Technologies (PTI-QTEP+) is also acknowledged.
\end{acknowledgments}

\clearpage

\appendix

\section{Dependence on $t$}
\label{appA}

Here we discuss the quantitative effect of $t$ on the low-energy eigenstates of the system.

\begin{figure}[htbp]
    \centering
    \includegraphics[width=1. \columnwidth]{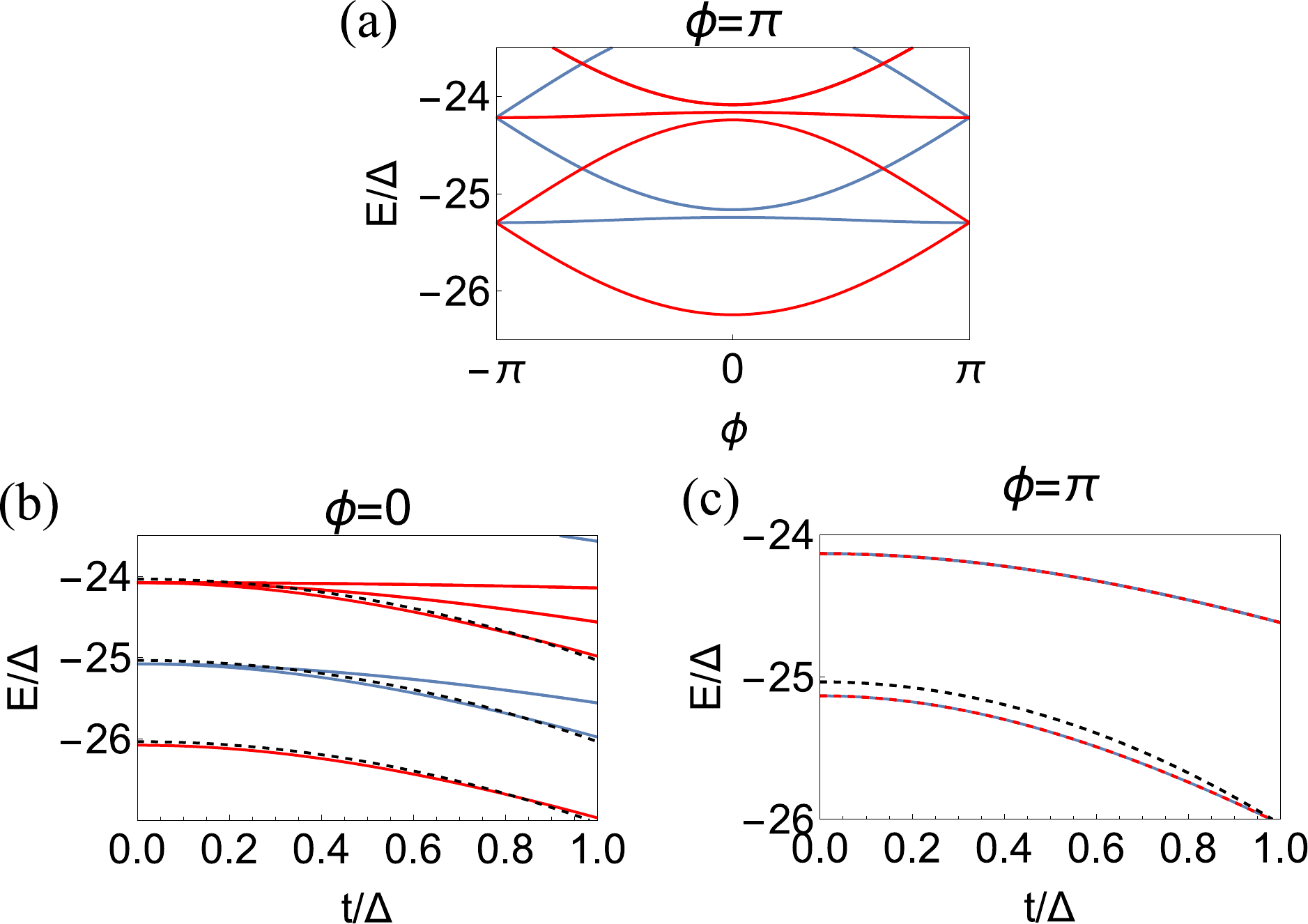}
    \caption{
    Quantitative effects of parameter $t$ on the spectra. All other parameters are the same as in Fig.~\ref{fig:2} of main text.
    (a) Low-energy spectra, singlet (red, $S=0$) and doublet (blue, $S=1/2$), for $t=0.4\Delta$.
    (b, c) Energies vs. $t$ for (b) $\phi=0$ and (c) $\phi=\pi$.
    Red: singlets, blue: doublets.
    Dashed black lines: perturbative corrections, Eq.~\eqref{eq:first_order}, with $\propto -t^2$ terms.
    }
    \label{fig:t_dependence}
\end{figure}

Fig.~\ref{fig:t_dependence}(a) shows the low-energy spectra for the same parameters as Fig.~\ref{fig:2}(c), but for much larger $t/\Delta = 0.4$. 
The main effect of larger $t$ is the splitting of the doublet states (blue lines) at $\phi = 0$. This stabilizes the lower energy state that has the expected symmetry properties that match those of the full problem.

In the language of the Hamiltonian in Eq.~\eqref{eq:SC}, $t$ energetically favours the formation of a singlet between $a(0)$ and $a(l)$, i.e., the state $[ a(0)^\dagger_\downarrow a(l)^\dagger_\uparrow - a(0)^\dagger_\uparrow a(l)^\dagger_\downarrow ] |0 \rangle$.
In the doublet manifold, the presence of a spin-carrying quasiparticle on either $a(0)$ or $a(l)$ hinders the formation of this singlet, so the energy lowering due to singlet formation is only possible if the quasiparticle occupies $b(l)$.
This leads to the splitting of the degeneracy at $\phi = 0$.

The energy gained with the inter-site singlet formation is $\sim t^2$. This is confirmed in Fig.~\ref{fig:t_dependence}(b, c), which shows the $t$ depdencene of the spectrum. The dashed black lines are perturbative corrections, Eq.~\eqref{eq:first_order}, to which we added $-t^2$ terms.

\vfill\eject

\section{Dependence on $\eta$}
\label{appB}

\begin{figure}[htbp]
    \centering
    \includegraphics[width=1. \columnwidth]{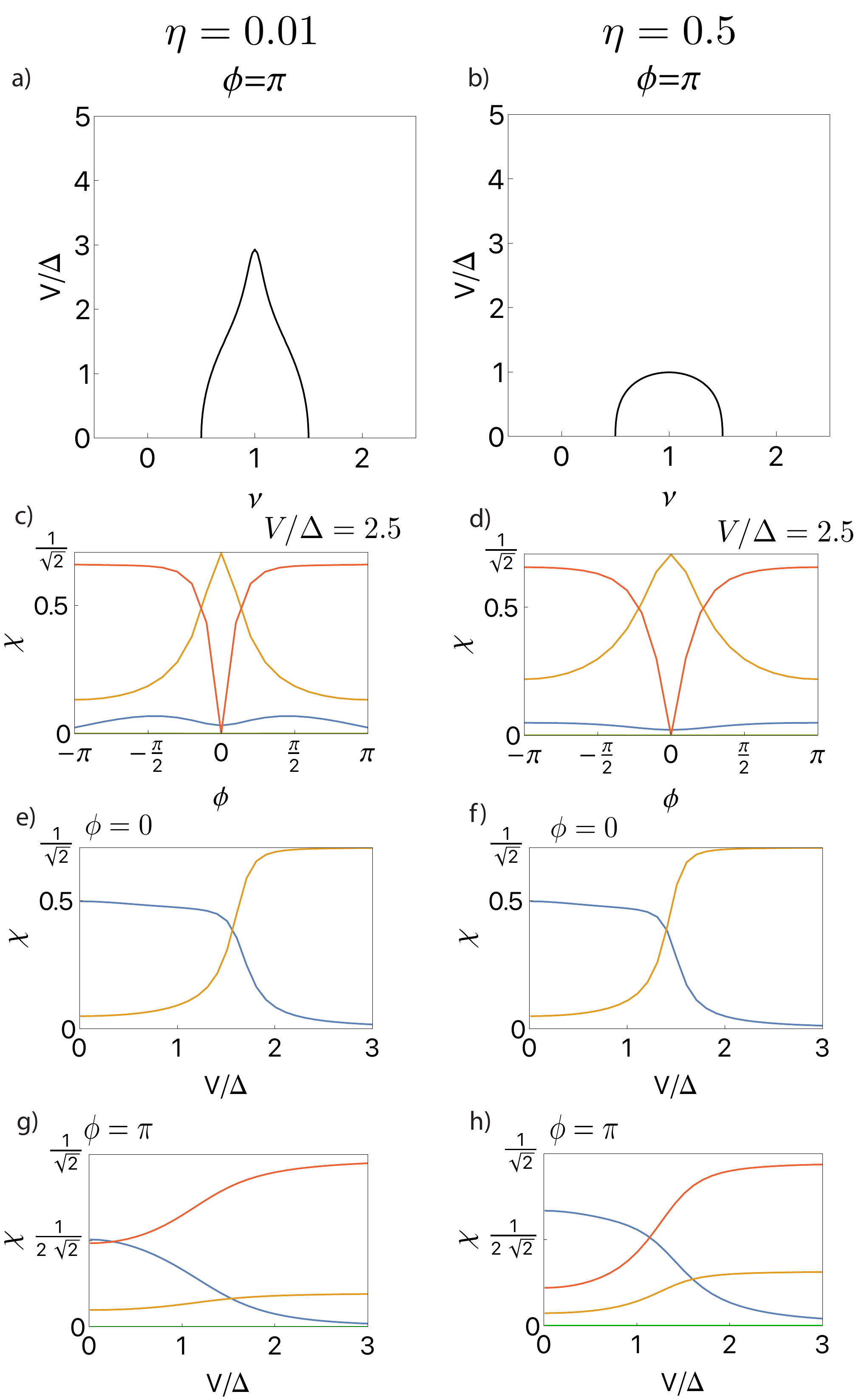}
    \caption{
    Phase diagrams and addition amplitudes for the left-right asymmetry parameter $\eta=0.01$ (left column) and $\eta=0.5$ (right column).
    (a,b) Phase diagrams at $\phi=\pi$. Parameters are $U=10\Delta$, $t=0.2\Delta$.
    (c,d) Addition amplitudes $\chi$ vs $\phi$.
    (e-h) Addition amplitudes $\chi$ vs $V$.
    }
    \label{fig:eta_dependence}
\end{figure}

In Fig.~\ref{fig:eta_dependence} we plot the phase diagrams and the addition amplitudes for two non-zero values of the left-right asymmetry parameter, a very small value $\eta=0.01$ and a large value $\eta=0.5$. These results should be compared to Fig.~\ref{fig:1}(d) and to Fig.~\ref{fig:screening}(b,c,d) in the main text. The main effects of the asymmetry are the disappearance of the doublet chimney at $\phi=\pi$ and the different composition of the doublet wavefunction at $\phi=\pi$. Both can be explained by the mapping of the left-right asymmetric problems at $\phi=\pi$ to the symmetric problem at some effective value of $\phi$ away from $\pi$ \cite{kadlecova2017}.
For $\eta=0.01$ there is still a visible protrusion of the doublet phase in the phase diagram for $\phi=\pi$, but it is no longer perceptible at $\eta=0.05$ (not shown). The doublet chimney therefore requires a rather high degree of left-right symmetry in the system.

\clearpage

\bibliography{bibl}

\begin{thebibliography}{49}%
\makeatletter
\providecommand \@ifxundefined [1]{%
 \@ifx{#1\undefined}
}%
\providecommand \@ifnum [1]{%
 \ifnum #1\expandafter \@firstoftwo
 \else \expandafter \@secondoftwo
 \fi
}%
\providecommand \@ifx [1]{%
 \ifx #1\expandafter \@firstoftwo
 \else \expandafter \@secondoftwo
 \fi
}%
\providecommand \natexlab [1]{#1}%
\providecommand \enquote  [1]{``#1''}%
\providecommand \bibnamefont  [1]{#1}%
\providecommand \bibfnamefont [1]{#1}%
\providecommand \citenamefont [1]{#1}%
\providecommand \href@noop [0]{\@secondoftwo}%
\providecommand \href [0]{\begingroup \@sanitize@url \@href}%
\providecommand \@href[1]{\@@startlink{#1}\@@href}%
\providecommand \@@href[1]{\endgroup#1\@@endlink}%
\providecommand \@sanitize@url [0]{\catcode `\\12\catcode `\$12\catcode
  `\&12\catcode `\#12\catcode `\^12\catcode `\_12\catcode `\%12\relax}%
\providecommand \@@startlink[1]{}%
\providecommand \@@endlink[0]{}%
\providecommand \url  [0]{\begingroup\@sanitize@url \@url }%
\providecommand \@url [1]{\endgroup\@href {#1}{\urlprefix }}%
\providecommand \urlprefix  [0]{URL }%
\providecommand \Eprint [0]{\href }%
\providecommand \doibase [0]{https://doi.org/}%
\providecommand \selectlanguage [0]{\@gobble}%
\providecommand \bibinfo  [0]{\@secondoftwo}%
\providecommand \bibfield  [0]{\@secondoftwo}%
\providecommand \translation [1]{[#1]}%
\providecommand \BibitemOpen [0]{}%
\providecommand \bibitemStop [0]{}%
\providecommand \bibitemNoStop [0]{.\EOS\space}%
\providecommand \EOS [0]{\spacefactor3000\relax}%
\providecommand \BibitemShut  [1]{\csname bibitem#1\endcsname}%
\let\auto@bib@innerbib\@empty
\bibitem [{\citenamefont {Devoret}\ and\ \citenamefont
  {Schoelkopf}(2013)}]{Devoret-Schoelkopf}%
  \BibitemOpen
  \bibfield  {author} {\bibinfo {author} {\bibfnamefont {M.~H.}\ \bibnamefont
  {Devoret}}\ and\ \bibinfo {author} {\bibfnamefont {R.~J.}\ \bibnamefont
  {Schoelkopf}},\ }\href {https://doi.org/10.1126/science.1231930} {\bibfield
  {journal} {\bibinfo  {journal} {Science}\ }\textbf {\bibinfo {volume}
  {339}},\ \bibinfo {pages} {1169} (\bibinfo {year} {2013})}\BibitemShut
  {NoStop}%
\bibitem [{\citenamefont {Aguado}(2020)}]{Aguado:APL20}%
  \BibitemOpen
  \bibfield  {author} {\bibinfo {author} {\bibfnamefont {R.}~\bibnamefont
  {Aguado}},\ }\href {https://doi.org/10.1063/5.0024124} {\bibfield  {journal}
  {\bibinfo  {journal} {Applied Physics Letters}\ }\textbf {\bibinfo {volume}
  {117}},\ \bibinfo {pages} {240501} (\bibinfo {year} {2020})}\BibitemShut
  {NoStop}%
\bibitem [{\citenamefont {Larsen}\ \emph {et~al.}(2020)\citenamefont {Larsen},
  \citenamefont {Gershenson}, \citenamefont {Casparis}, \citenamefont
  {Kringh\o{}j}, \citenamefont {Pearson}, \citenamefont {McNeil}, \citenamefont
  {Kuemmeth}, \citenamefont {Krogstrup}, \citenamefont {Petersson},\ and\
  \citenamefont {Marcus}}]{Larsen:2020}%
  \BibitemOpen
  \bibfield  {author} {\bibinfo {author} {\bibfnamefont {T.~W.}\ \bibnamefont
  {Larsen}}, \bibinfo {author} {\bibfnamefont {M.~E.}\ \bibnamefont
  {Gershenson}}, \bibinfo {author} {\bibfnamefont {L.}~\bibnamefont
  {Casparis}}, \bibinfo {author} {\bibfnamefont {A.}~\bibnamefont
  {Kringh\o{}j}}, \bibinfo {author} {\bibfnamefont {N.~J.}\ \bibnamefont
  {Pearson}}, \bibinfo {author} {\bibfnamefont {R.~P.~G.}\ \bibnamefont
  {McNeil}}, \bibinfo {author} {\bibfnamefont {F.}~\bibnamefont {Kuemmeth}},
  \bibinfo {author} {\bibfnamefont {P.}~\bibnamefont {Krogstrup}}, \bibinfo
  {author} {\bibfnamefont {K.~D.}\ \bibnamefont {Petersson}},\ and\ \bibinfo
  {author} {\bibfnamefont {C.~M.}\ \bibnamefont {Marcus}},\ }\href
  {https://doi.org/10.1103/PhysRevLett.125.056801} {\bibfield  {journal}
  {\bibinfo  {journal} {Phys. Rev. Lett.}\ }\textbf {\bibinfo {volume} {125}},\
  \bibinfo {pages} {056801} (\bibinfo {year} {2020})}\BibitemShut {NoStop}%
\bibitem [{\citenamefont {Zazunov}\ \emph {et~al.}(2003)\citenamefont
  {Zazunov}, \citenamefont {Shumeiko}, \citenamefont {Bratus'}, \citenamefont
  {Lantz},\ and\ \citenamefont {Wendin}}]{Zazunov2003}%
  \BibitemOpen
  \bibfield  {author} {\bibinfo {author} {\bibfnamefont {A.}~\bibnamefont
  {Zazunov}}, \bibinfo {author} {\bibfnamefont {V.~S.}\ \bibnamefont
  {Shumeiko}}, \bibinfo {author} {\bibfnamefont {E.~N.}\ \bibnamefont
  {Bratus'}}, \bibinfo {author} {\bibfnamefont {J.}~\bibnamefont {Lantz}},\
  and\ \bibinfo {author} {\bibfnamefont {G.}~\bibnamefont {Wendin}},\
  }\bibfield  {journal} {\bibinfo  {journal} {Physical Review Letters}\
  }\textbf {\bibinfo {volume} {90}},\ \href
  {https://doi.org/10.1103/physrevlett.90.087003}
  {10.1103/physrevlett.90.087003} (\bibinfo {year} {2003})\BibitemShut
  {NoStop}%
\bibitem [{\citenamefont {Chtchelkatchev}\ and\ \citenamefont
  {Nazarov}(2003)}]{Chtchelkatchev:2003}%
  \BibitemOpen
  \bibfield  {author} {\bibinfo {author} {\bibfnamefont {N.~M.}\ \bibnamefont
  {Chtchelkatchev}}\ and\ \bibinfo {author} {\bibfnamefont {Y.~V.}\
  \bibnamefont {Nazarov}},\ }\href
  {https://doi.org/10.1103/PhysRevLett.90.226806} {\bibfield  {journal}
  {\bibinfo  {journal} {Phys. Rev. Lett.}\ }\textbf {\bibinfo {volume} {90}},\
  \bibinfo {pages} {226806} (\bibinfo {year} {2003})}\BibitemShut {NoStop}%
\bibitem [{\citenamefont {Padurariu}\ and\ \citenamefont
  {Nazarov}(2010)}]{Padurari:2010}%
  \BibitemOpen
  \bibfield  {author} {\bibinfo {author} {\bibfnamefont {C.}~\bibnamefont
  {Padurariu}}\ and\ \bibinfo {author} {\bibfnamefont {Y.~V.}\ \bibnamefont
  {Nazarov}},\ }\href {https://doi.org/10.1103/PhysRevB.81.144519} {\bibfield
  {journal} {\bibinfo  {journal} {Phys. Rev. B}\ }\textbf {\bibinfo {volume}
  {81}},\ \bibinfo {pages} {144519} (\bibinfo {year} {2010})}\BibitemShut
  {NoStop}%
\bibitem [{\citenamefont {Hays}\ \emph {et~al.}(2021)\citenamefont {Hays},
  \citenamefont {Fatemi}, \citenamefont {Bouman}, \citenamefont {Cerrillo},
  \citenamefont {Diamond}, \citenamefont {Serniak}, \citenamefont {Connolly},
  \citenamefont {Krogstrup}, \citenamefont {Nyg{\aa}rd}, \citenamefont
  {Yeyati}, \citenamefont {Geresdi},\ and\ \citenamefont {Devoret}}]{Hays2021}%
  \BibitemOpen
  \bibfield  {author} {\bibinfo {author} {\bibfnamefont {M.}~\bibnamefont
  {Hays}}, \bibinfo {author} {\bibfnamefont {V.}~\bibnamefont {Fatemi}},
  \bibinfo {author} {\bibfnamefont {D.}~\bibnamefont {Bouman}}, \bibinfo
  {author} {\bibfnamefont {J.}~\bibnamefont {Cerrillo}}, \bibinfo {author}
  {\bibfnamefont {S.}~\bibnamefont {Diamond}}, \bibinfo {author} {\bibfnamefont
  {K.}~\bibnamefont {Serniak}}, \bibinfo {author} {\bibfnamefont
  {T.}~\bibnamefont {Connolly}}, \bibinfo {author} {\bibfnamefont
  {P.}~\bibnamefont {Krogstrup}}, \bibinfo {author} {\bibfnamefont
  {J.}~\bibnamefont {Nyg{\aa}rd}}, \bibinfo {author} {\bibfnamefont {A.~L.}\
  \bibnamefont {Yeyati}}, \bibinfo {author} {\bibfnamefont {A.}~\bibnamefont
  {Geresdi}},\ and\ \bibinfo {author} {\bibfnamefont {M.~H.}\ \bibnamefont
  {Devoret}},\ }\href {https://doi.org/10.1126/science.abf0345} {\bibfield
  {journal} {\bibinfo  {journal} {Science}\ }\textbf {\bibinfo {volume}
  {373}},\ \bibinfo {pages} {430} (\bibinfo {year} {2021})}\BibitemShut
  {NoStop}%
\bibitem [{\citenamefont {Bargerbos}\ \emph {et~al.}(2022)\citenamefont
  {Bargerbos}, \citenamefont {Pita-Vidal}, \citenamefont {{\v{Z}}itko},
  \citenamefont {{\'{A}}vila}, \citenamefont {Splitthoff}, \citenamefont
  {Gr\"{u}nhaupt}, \citenamefont {Wesdorp}, \citenamefont {Andersen},
  \citenamefont {Liu}, \citenamefont {Kouwenhoven}, \citenamefont {Aguado},
  \citenamefont {Kou},\ and\ \citenamefont {van Heck}}]{Bargerbos2022}%
  \BibitemOpen
  \bibfield  {author} {\bibinfo {author} {\bibfnamefont {A.}~\bibnamefont
  {Bargerbos}}, \bibinfo {author} {\bibfnamefont {M.}~\bibnamefont
  {Pita-Vidal}}, \bibinfo {author} {\bibfnamefont {R.}~\bibnamefont
  {{\v{Z}}itko}}, \bibinfo {author} {\bibfnamefont {J.}~\bibnamefont
  {{\'{A}}vila}}, \bibinfo {author} {\bibfnamefont {L.~J.}\ \bibnamefont
  {Splitthoff}}, \bibinfo {author} {\bibfnamefont {L.}~\bibnamefont
  {Gr\"{u}nhaupt}}, \bibinfo {author} {\bibfnamefont {J.~J.}\ \bibnamefont
  {Wesdorp}}, \bibinfo {author} {\bibfnamefont {C.~K.}\ \bibnamefont
  {Andersen}}, \bibinfo {author} {\bibfnamefont {Y.}~\bibnamefont {Liu}},
  \bibinfo {author} {\bibfnamefont {L.~P.}\ \bibnamefont {Kouwenhoven}},
  \bibinfo {author} {\bibfnamefont {R.}~\bibnamefont {Aguado}}, \bibinfo
  {author} {\bibfnamefont {A.}~\bibnamefont {Kou}},\ and\ \bibinfo {author}
  {\bibfnamefont {B.}~\bibnamefont {van Heck}},\ }\bibfield  {journal}
  {\bibinfo  {journal} {{PRX} Quantum}\ }\textbf {\bibinfo {volume} {3}},\
  \href {https://doi.org/10.1103/prxquantum.3.030311}
  {10.1103/prxquantum.3.030311} (\bibinfo {year} {2022})\BibitemShut {NoStop}%
\bibitem [{\citenamefont {Bargerbos}\ \emph {et~al.}(2023)\citenamefont
  {Bargerbos}, \citenamefont {Pita-Vidal}, \citenamefont {Žitko},
  \citenamefont {Splitthoff}, \citenamefont {Grünhaupt}, \citenamefont
  {Wesdorp}, \citenamefont {Liu}, \citenamefont {Kouwenhoven}, \citenamefont
  {Aguado}, \citenamefont {Andersen}, \citenamefont {Kou},\ and\ \citenamefont
  {van Heck}}]{Bargerbos:22b}%
  \BibitemOpen
  \bibfield  {author} {\bibinfo {author} {\bibfnamefont {A.}~\bibnamefont
  {Bargerbos}}, \bibinfo {author} {\bibfnamefont {M.}~\bibnamefont
  {Pita-Vidal}}, \bibinfo {author} {\bibfnamefont {R.}~\bibnamefont {Žitko}},
  \bibinfo {author} {\bibfnamefont {L.~J.}\ \bibnamefont {Splitthoff}},
  \bibinfo {author} {\bibfnamefont {L.}~\bibnamefont {Grünhaupt}}, \bibinfo
  {author} {\bibfnamefont {J.~J.}\ \bibnamefont {Wesdorp}}, \bibinfo {author}
  {\bibfnamefont {Y.}~\bibnamefont {Liu}}, \bibinfo {author} {\bibfnamefont
  {L.~P.}\ \bibnamefont {Kouwenhoven}}, \bibinfo {author} {\bibfnamefont
  {R.}~\bibnamefont {Aguado}}, \bibinfo {author} {\bibfnamefont {C.~K.}\
  \bibnamefont {Andersen}}, \bibinfo {author} {\bibfnamefont {A.}~\bibnamefont
  {Kou}},\ and\ \bibinfo {author} {\bibfnamefont {B.}~\bibnamefont {van
  Heck}},\ }\href@noop {} {\bibfield  {journal} {\bibinfo  {journal} {Phys.
  Rev. Lett.}\ }\textbf {\bibinfo {volume} {131}},\ \bibinfo {pages} {097001}
  (\bibinfo {year} {2023})}\BibitemShut {NoStop}%
\bibitem [{\citenamefont {Pita-Vidal}\ \emph {et~al.}(2023)\citenamefont
  {Pita-Vidal}, \citenamefont {Bargerbos}, \citenamefont {{\v{Z}}itko},
  \citenamefont {Splitthoff}, \citenamefont {Gr\"{u}nhaupt}, \citenamefont
  {Wesdorp}, \citenamefont {Liu}, \citenamefont {Kouwenhoven}, \citenamefont
  {Aguado}, \citenamefont {van Heck}, \citenamefont {Kou},\ and\ \citenamefont
  {Andersen}}]{Pita-Vidal:2022}%
  \BibitemOpen
  \bibfield  {author} {\bibinfo {author} {\bibfnamefont {M.}~\bibnamefont
  {Pita-Vidal}}, \bibinfo {author} {\bibfnamefont {A.}~\bibnamefont
  {Bargerbos}}, \bibinfo {author} {\bibfnamefont {R.}~\bibnamefont
  {{\v{Z}}itko}}, \bibinfo {author} {\bibfnamefont {L.~J.}\ \bibnamefont
  {Splitthoff}}, \bibinfo {author} {\bibfnamefont {L.}~\bibnamefont
  {Gr\"{u}nhaupt}}, \bibinfo {author} {\bibfnamefont {J.~J.}\ \bibnamefont
  {Wesdorp}}, \bibinfo {author} {\bibfnamefont {Y.}~\bibnamefont {Liu}},
  \bibinfo {author} {\bibfnamefont {L.~P.}\ \bibnamefont {Kouwenhoven}},
  \bibinfo {author} {\bibfnamefont {R.}~\bibnamefont {Aguado}}, \bibinfo
  {author} {\bibfnamefont {B.}~\bibnamefont {van Heck}}, \bibinfo {author}
  {\bibfnamefont {A.}~\bibnamefont {Kou}},\ and\ \bibinfo {author}
  {\bibfnamefont {C.~K.}\ \bibnamefont {Andersen}},\ }\bibfield  {journal}
  {\bibinfo  {journal} {Nature Physics}\ }\href
  {https://doi.org/10.1038/s41567-023-02071-x} {10.1038/s41567-023-02071-x}
  (\bibinfo {year} {2023})\BibitemShut {NoStop}%
\bibitem [{\citenamefont {Kir{\v{s}}anskas}\ \emph {et~al.}(2015)\citenamefont
  {Kir{\v{s}}anskas}, \citenamefont {Goldstein}, \citenamefont {Flensberg},
  \citenamefont {Glazman},\ and\ \citenamefont {Paaske}}]{Kirsanskas2015}%
  \BibitemOpen
  \bibfield  {author} {\bibinfo {author} {\bibfnamefont {G.}~\bibnamefont
  {Kir{\v{s}}anskas}}, \bibinfo {author} {\bibfnamefont {M.}~\bibnamefont
  {Goldstein}}, \bibinfo {author} {\bibfnamefont {K.}~\bibnamefont
  {Flensberg}}, \bibinfo {author} {\bibfnamefont {L.~I.}\ \bibnamefont
  {Glazman}},\ and\ \bibinfo {author} {\bibfnamefont {J.}~\bibnamefont
  {Paaske}},\ }\href {https://doi.org/10.1103/physrevb.92.235422} {\bibfield
  {journal} {\bibinfo  {journal} {Physical Review B}\ }\textbf {\bibinfo
  {volume} {92}},\ \bibinfo {pages} {235422} (\bibinfo {year}
  {2015})}\BibitemShut {NoStop}%
\bibitem [{\citenamefont {Meden}(2019)}]{Meden2019}%
  \BibitemOpen
  \bibfield  {author} {\bibinfo {author} {\bibfnamefont {V.}~\bibnamefont
  {Meden}},\ }\href {https://doi.org/10.1088/1361-648x/aafd6a} {\bibfield
  {journal} {\bibinfo  {journal} {Journal of Physics: Condensed Matter}\
  }\textbf {\bibinfo {volume} {31}},\ \bibinfo {pages} {163001} (\bibinfo
  {year} {2019})}\BibitemShut {NoStop}%
\bibitem [{\citenamefont {Affleck}\ \emph {et~al.}(2000)\citenamefont
  {Affleck}, \citenamefont {Caux},\ and\ \citenamefont
  {Zagoskin}}]{affleck2000}%
  \BibitemOpen
  \bibfield  {author} {\bibinfo {author} {\bibfnamefont {I.}~\bibnamefont
  {Affleck}}, \bibinfo {author} {\bibfnamefont {J.-S.}\ \bibnamefont {Caux}},\
  and\ \bibinfo {author} {\bibfnamefont {A.~M.}\ \bibnamefont {Zagoskin}},\
  }\href {https://doi.org/10.1103/physrevb.62.1433} {\bibfield  {journal}
  {\bibinfo  {journal} {Physical Review B}\ }\textbf {\bibinfo {volume} {62}},\
  \bibinfo {pages} {1433} (\bibinfo {year} {2000})}\BibitemShut {NoStop}%
\bibitem [{\citenamefont {Vecino}\ \emph {et~al.}(2003)\citenamefont {Vecino},
  \citenamefont {Mart{\'{\i}}n-Rodero},\ and\ \citenamefont
  {Yeyati}}]{Vecino2003}%
  \BibitemOpen
  \bibfield  {author} {\bibinfo {author} {\bibfnamefont {E.}~\bibnamefont
  {Vecino}}, \bibinfo {author} {\bibfnamefont {A.}~\bibnamefont
  {Mart{\'{\i}}n-Rodero}},\ and\ \bibinfo {author} {\bibfnamefont {A.~L.}\
  \bibnamefont {Yeyati}},\ }\bibfield  {journal} {\bibinfo  {journal} {Physical
  Review B}\ }\textbf {\bibinfo {volume} {68}},\ \href
  {https://doi.org/10.1103/physrevb.68.035105} {10.1103/physrevb.68.035105}
  (\bibinfo {year} {2003})\BibitemShut {NoStop}%
\bibitem [{\citenamefont {Bergeret}\ \emph {et~al.}(2007)\citenamefont
  {Bergeret}, \citenamefont {Yeyati},\ and\ \citenamefont
  {Mart{\'{\i}}n-Rodero}}]{bergeret2007}%
  \BibitemOpen
  \bibfield  {author} {\bibinfo {author} {\bibfnamefont {F.~S.}\ \bibnamefont
  {Bergeret}}, \bibinfo {author} {\bibfnamefont {A.~L.}\ \bibnamefont
  {Yeyati}},\ and\ \bibinfo {author} {\bibfnamefont {A.}~\bibnamefont
  {Mart{\'{\i}}n-Rodero}},\ }\bibfield  {journal} {\bibinfo  {journal}
  {Physical Review B}\ }\textbf {\bibinfo {volume} {76}},\ \href
  {https://doi.org/10.1103/physrevb.76.174510} {10.1103/physrevb.76.174510}
  (\bibinfo {year} {2007})\BibitemShut {NoStop}%
\bibitem [{\citenamefont {Grove-Rasmussen}\ \emph {et~al.}(2018)\citenamefont
  {Grove-Rasmussen}, \citenamefont {Steffensen}, \citenamefont {Jellinggaard},
  \citenamefont {Madsen}, \citenamefont {{\v{Z}}itko}, \citenamefont {Paaske},\
  and\ \citenamefont {Nyg{\aa}rd}}]{GroveRasmussen2018}%
  \BibitemOpen
  \bibfield  {author} {\bibinfo {author} {\bibfnamefont {K.}~\bibnamefont
  {Grove-Rasmussen}}, \bibinfo {author} {\bibfnamefont {G.}~\bibnamefont
  {Steffensen}}, \bibinfo {author} {\bibfnamefont {A.}~\bibnamefont
  {Jellinggaard}}, \bibinfo {author} {\bibfnamefont {M.~H.}\ \bibnamefont
  {Madsen}}, \bibinfo {author} {\bibfnamefont {R.}~\bibnamefont {{\v{Z}}itko}},
  \bibinfo {author} {\bibfnamefont {J.}~\bibnamefont {Paaske}},\ and\ \bibinfo
  {author} {\bibfnamefont {J.}~\bibnamefont {Nyg{\aa}rd}},\ }\bibfield
  {journal} {\bibinfo  {journal} {Nature Communications}\ }\textbf {\bibinfo
  {volume} {9}},\ \href {https://doi.org/10.1038/s41467-018-04683-x}
  {10.1038/s41467-018-04683-x} (\bibinfo {year} {2018})\BibitemShut {NoStop}%
\bibitem [{\citenamefont {Saldaña}\ \emph {et~al.}(2022)\citenamefont
  {Saldaña}, \citenamefont {Vekris}, \citenamefont {Pavešič}, \citenamefont
  {Žitko}, \citenamefont {Grove-Rasmussen},\ and\ \citenamefont
  {Nygård}}]{twoBogoliubov}%
  \BibitemOpen
  \bibfield  {author} {\bibinfo {author} {\bibfnamefont {J.~C.~E.}\
  \bibnamefont {Saldaña}}, \bibinfo {author} {\bibfnamefont {A.}~\bibnamefont
  {Vekris}}, \bibinfo {author} {\bibfnamefont {L.}~\bibnamefont {Pavešič}},
  \bibinfo {author} {\bibfnamefont {R.}~\bibnamefont {Žitko}}, \bibinfo
  {author} {\bibfnamefont {K.}~\bibnamefont {Grove-Rasmussen}},\ and\ \bibinfo
  {author} {\bibfnamefont {J.}~\bibnamefont {Nygård}},\ }\href
  {https://doi.org/10.48550/ARXIV.2203.00104} {\bibinfo {title} {{Correlation
  between two distant quasiparticles in separate superconducting islands
  mediated by a single spin}}} (\bibinfo {year} {2022})\BibitemShut {NoStop}%
\bibitem [{\citenamefont {Žonda}\ \emph {et~al.}(2023)\citenamefont {Žonda},
  \citenamefont {Zalom}, \citenamefont {Novotný}, \citenamefont {Loukeris},
  \citenamefont {B\"{a}tge},\ and\ \citenamefont {Pokorný}}]{Zonda2022}%
  \BibitemOpen
  \bibfield  {author} {\bibinfo {author} {\bibfnamefont {M.}~\bibnamefont
  {Žonda}}, \bibinfo {author} {\bibfnamefont {P.}~\bibnamefont {Zalom}},
  \bibinfo {author} {\bibfnamefont {T.}~\bibnamefont {Novotný}}, \bibinfo
  {author} {\bibfnamefont {G.}~\bibnamefont {Loukeris}}, \bibinfo {author}
  {\bibfnamefont {J.}~\bibnamefont {B\"{a}tge}},\ and\ \bibinfo {author}
  {\bibfnamefont {V.}~\bibnamefont {Pokorný}},\ }\href@noop {} {\bibinfo
  {title} {Generalized atomic limit of a double quantum dot coupled to
  superconducting leads}} (\bibinfo {year} {2023})\BibitemShut {NoStop}%
\bibitem [{\citenamefont {Hermansen}\ \emph {et~al.}(2022)\citenamefont
  {Hermansen}, \citenamefont {Yeyati},\ and\ \citenamefont
  {Paaske}}]{Hermansen2022}%
  \BibitemOpen
  \bibfield  {author} {\bibinfo {author} {\bibfnamefont {C.}~\bibnamefont
  {Hermansen}}, \bibinfo {author} {\bibfnamefont {A.~L.}\ \bibnamefont
  {Yeyati}},\ and\ \bibinfo {author} {\bibfnamefont {J.}~\bibnamefont
  {Paaske}},\ }\bibfield  {journal} {\bibinfo  {journal} {Physical Review B}\
  }\textbf {\bibinfo {volume} {105}},\ \href
  {https://doi.org/10.1103/physrevb.105.054503} {10.1103/physrevb.105.054503}
  (\bibinfo {year} {2022})\BibitemShut {NoStop}%
\bibitem [{\citenamefont {Schmid}\ \emph {et~al.}(2022)\citenamefont {Schmid},
  \citenamefont {Steiner}, \citenamefont {Franke},\ and\ \citenamefont {von
  Oppen}}]{Schmid2022}%
  \BibitemOpen
  \bibfield  {author} {\bibinfo {author} {\bibfnamefont {H.}~\bibnamefont
  {Schmid}}, \bibinfo {author} {\bibfnamefont {J.~F.}\ \bibnamefont {Steiner}},
  \bibinfo {author} {\bibfnamefont {K.~J.}\ \bibnamefont {Franke}},\ and\
  \bibinfo {author} {\bibfnamefont {F.}~\bibnamefont {von Oppen}},\ }\bibfield
  {journal} {\bibinfo  {journal} {Physical Review B}\ }\textbf {\bibinfo
  {volume} {105}},\ \href {https://doi.org/10.1103/physrevb.105.235406}
  {10.1103/physrevb.105.235406} (\bibinfo {year} {2022})\BibitemShut {NoStop}%
\bibitem [{\citenamefont {Moca}\ \emph {et~al.}(2021)\citenamefont {Moca},
  \citenamefont {Weymann}, \citenamefont {Werner},\ and\ \citenamefont
  {Zar{\'{a}}nd}}]{Moca2021}%
  \BibitemOpen
  \bibfield  {author} {\bibinfo {author} {\bibfnamefont {C.~P.}\ \bibnamefont
  {Moca}}, \bibinfo {author} {\bibfnamefont {I.}~\bibnamefont {Weymann}},
  \bibinfo {author} {\bibfnamefont {M.~A.}\ \bibnamefont {Werner}},\ and\
  \bibinfo {author} {\bibfnamefont {G.}~\bibnamefont {Zar{\'{a}}nd}},\
  }\bibfield  {journal} {\bibinfo  {journal} {Physical Review Letters}\
  }\textbf {\bibinfo {volume} {127}},\ \href
  {https://doi.org/10.1103/physrevlett.127.186804}
  {10.1103/physrevlett.127.186804} (\bibinfo {year} {2021})\BibitemShut
  {NoStop}%
\bibitem [{\citenamefont {Pave\v{s}i\'c}\ \emph {et~al.}(2022)\citenamefont
  {Pave\v{s}i\'c}, \citenamefont {Pita-Vidal}, \citenamefont {Bargerbos},\ and\
  \citenamefont {\v{Z}itko}}]{knight}%
  \BibitemOpen
  \bibfield  {author} {\bibinfo {author} {\bibfnamefont {L.}~\bibnamefont
  {Pave\v{s}i\'c}}, \bibinfo {author} {\bibfnamefont {M.}~\bibnamefont
  {Pita-Vidal}}, \bibinfo {author} {\bibfnamefont {A.}~\bibnamefont
  {Bargerbos}},\ and\ \bibinfo {author} {\bibfnamefont {R.}~\bibnamefont
  {\v{Z}itko}},\ }\href@noop {} {\bibinfo {title} {Impurity {Knight} shift in
  quantum dot {Josephson} junctions}},\ \bibinfo {howpublished}
  {arXiv:2212.07185} (\bibinfo {year} {2022})\BibitemShut {NoStop}%
\bibitem [{\citenamefont {Shiba}\ and\ \citenamefont {Soda}(1969)}]{Shiba1969}%
  \BibitemOpen
  \bibfield  {author} {\bibinfo {author} {\bibfnamefont {H.}~\bibnamefont
  {Shiba}}\ and\ \bibinfo {author} {\bibfnamefont {T.}~\bibnamefont {Soda}},\
  }\href {https://doi.org/10.1143/ptp.41.25} {\bibfield  {journal} {\bibinfo
  {journal} {Progress of Theoretical Physics}\ }\textbf {\bibinfo {volume}
  {41}},\ \bibinfo {pages} {25} (\bibinfo {year} {1969})}\BibitemShut {NoStop}%
\bibitem [{\citenamefont {Matveev}\ \emph {et~al.}(1993)\citenamefont
  {Matveev}, \citenamefont {Gisself\"{a}lt}, \citenamefont {Glazman},
  \citenamefont {Jonson},\ and\ \citenamefont {Shekhter}}]{Matveev1993}%
  \BibitemOpen
  \bibfield  {author} {\bibinfo {author} {\bibfnamefont {K.~A.}\ \bibnamefont
  {Matveev}}, \bibinfo {author} {\bibfnamefont {M.}~\bibnamefont
  {Gisself\"{a}lt}}, \bibinfo {author} {\bibfnamefont {L.~I.}\ \bibnamefont
  {Glazman}}, \bibinfo {author} {\bibfnamefont {M.}~\bibnamefont {Jonson}},\
  and\ \bibinfo {author} {\bibfnamefont {R.~I.}\ \bibnamefont {Shekhter}},\
  }\href {https://doi.org/10.1103/physrevlett.70.2940} {\bibfield  {journal}
  {\bibinfo  {journal} {Physical Review Letters}\ }\textbf {\bibinfo {volume}
  {70}},\ \bibinfo {pages} {2940} (\bibinfo {year} {1993})}\BibitemShut
  {NoStop}%
\bibitem [{\citenamefont {Rozhkov}\ and\ \citenamefont
  {Arovas}(1999)}]{Rozhkov1999}%
  \BibitemOpen
  \bibfield  {author} {\bibinfo {author} {\bibfnamefont {A.~V.}\ \bibnamefont
  {Rozhkov}}\ and\ \bibinfo {author} {\bibfnamefont {D.~P.}\ \bibnamefont
  {Arovas}},\ }\href {https://doi.org/10.1103/physrevlett.82.2788} {\bibfield
  {journal} {\bibinfo  {journal} {Physical Review Letters}\ }\textbf {\bibinfo
  {volume} {82}},\ \bibinfo {pages} {2788} (\bibinfo {year}
  {1999})}\BibitemShut {NoStop}%
\bibitem [{\citenamefont {Clerk}\ and\ \citenamefont
  {Ambegaokar}(2000)}]{Clerk2000}%
  \BibitemOpen
  \bibfield  {author} {\bibinfo {author} {\bibfnamefont {A.~A.}\ \bibnamefont
  {Clerk}}\ and\ \bibinfo {author} {\bibfnamefont {V.}~\bibnamefont
  {Ambegaokar}},\ }\href {https://doi.org/10.1103/physrevb.61.9109} {\bibfield
  {journal} {\bibinfo  {journal} {Physical Review B}\ }\textbf {\bibinfo
  {volume} {61}},\ \bibinfo {pages} {9109} (\bibinfo {year}
  {2000})}\BibitemShut {NoStop}%
\bibitem [{\citenamefont {Siano}\ and\ \citenamefont
  {Egger}(2004)}]{Siano2004}%
  \BibitemOpen
  \bibfield  {author} {\bibinfo {author} {\bibfnamefont {F.}~\bibnamefont
  {Siano}}\ and\ \bibinfo {author} {\bibfnamefont {R.}~\bibnamefont {Egger}},\
  }\bibfield  {journal} {\bibinfo  {journal} {Physical Review Letters}\
  }\textbf {\bibinfo {volume} {93}},\ \href
  {https://doi.org/10.1103/physrevlett.93.047002}
  {10.1103/physrevlett.93.047002} (\bibinfo {year} {2004})\BibitemShut
  {NoStop}%
\bibitem [{\citenamefont {Oguri}\ \emph {et~al.}(2004)\citenamefont {Oguri},
  \citenamefont {Tanaka},\ and\ \citenamefont {Hewson}}]{Oguri2004}%
  \BibitemOpen
  \bibfield  {author} {\bibinfo {author} {\bibfnamefont {A.}~\bibnamefont
  {Oguri}}, \bibinfo {author} {\bibfnamefont {Y.}~\bibnamefont {Tanaka}},\ and\
  \bibinfo {author} {\bibfnamefont {A.~C.}\ \bibnamefont {Hewson}},\ }\href
  {https://doi.org/10.1143/jpsj.73.2494} {\bibfield  {journal} {\bibinfo
  {journal} {Journal of the Physical Society of Japan}\ }\textbf {\bibinfo
  {volume} {73}},\ \bibinfo {pages} {2494} (\bibinfo {year}
  {2004})}\BibitemShut {NoStop}%
\bibitem [{\citenamefont {Choi}\ \emph {et~al.}(2004)\citenamefont {Choi},
  \citenamefont {Lee}, \citenamefont {Kang},\ and\ \citenamefont
  {Belzig}}]{Choi2004}%
  \BibitemOpen
  \bibfield  {author} {\bibinfo {author} {\bibfnamefont {M.-S.}\ \bibnamefont
  {Choi}}, \bibinfo {author} {\bibfnamefont {M.}~\bibnamefont {Lee}}, \bibinfo
  {author} {\bibfnamefont {K.}~\bibnamefont {Kang}},\ and\ \bibinfo {author}
  {\bibfnamefont {W.}~\bibnamefont {Belzig}},\ }\bibfield  {journal} {\bibinfo
  {journal} {Physical Review B}\ }\textbf {\bibinfo {volume} {70}},\ \href
  {https://doi.org/10.1103/physrevb.70.020502} {10.1103/physrevb.70.020502}
  (\bibinfo {year} {2004})\BibitemShut {NoStop}%
\bibitem [{\citenamefont {Yao}\ and\ \citenamefont {Shi}(2000)}]{Yao2000}%
  \BibitemOpen
  \bibfield  {author} {\bibinfo {author} {\bibfnamefont {D.}~\bibnamefont
  {Yao}}\ and\ \bibinfo {author} {\bibfnamefont {J.}~\bibnamefont {Shi}},\
  }\href {https://doi.org/10.1119/1.19419} {\bibfield  {journal} {\bibinfo
  {journal} {American Journal of Physics}\ }\textbf {\bibinfo {volume} {68}},\
  \bibinfo {pages} {278} (\bibinfo {year} {2000})}\BibitemShut {NoStop}%
\bibitem [{\citenamefont {Kadlecov{\'{a}}}\ \emph {et~al.}(2017)\citenamefont
  {Kadlecov{\'{a}}}, \citenamefont {{\v{Z}}onda},\ and\ \citenamefont
  {Novotn{\'{y}}}}]{kadlecova2017}%
  \BibitemOpen
  \bibfield  {author} {\bibinfo {author} {\bibfnamefont {A.}~\bibnamefont
  {Kadlecov{\'{a}}}}, \bibinfo {author} {\bibfnamefont {M.}~\bibnamefont
  {{\v{Z}}onda}},\ and\ \bibinfo {author} {\bibfnamefont {T.}~\bibnamefont
  {Novotn{\'{y}}}},\ }\bibfield  {journal} {\bibinfo  {journal} {Physical
  Review B}\ }\textbf {\bibinfo {volume} {95}},\ \href
  {https://doi.org/10.1103/physrevb.95.195114} {10.1103/physrevb.95.195114}
  (\bibinfo {year} {2017})\BibitemShut {NoStop}%
\bibitem [{\citenamefont {Combescot}(2022)}]{combescot}%
  \BibitemOpen
  \bibfield  {author} {\bibinfo {author} {\bibfnamefont {R.}~\bibnamefont
  {Combescot}},\ }\href@noop {} {\emph {\bibinfo {title} {Superconductivity: An
  introduction}}}\ (\bibinfo  {publisher} {Cambridge University Press},\
  \bibinfo {year} {2022})\BibitemShut {NoStop}%
\bibitem [{\citenamefont {\v{Z}itko}(2022)}]{rok_zitko_2022_5874832}%
  \BibitemOpen
  \bibfield  {author} {\bibinfo {author} {\bibfnamefont {R.}~\bibnamefont
  {\v{Z}itko}},\ }\href {https://doi.org/10.5281/zenodo.5874832} {\bibinfo
  {title} {{Josephson potentials for single impurity Anderson impurity in a
  junction between two superconductors}}},\ \bibinfo {howpublished} {Zenodo
  dataset, 10.5281/zenodo.5874832} (\bibinfo {year} {2022})\BibitemShut
  {NoStop}%
\bibitem [{\citenamefont {von Delft}\ and\ \citenamefont
  {Ralph}(2001)}]{vonDelft2001}%
  \BibitemOpen
  \bibfield  {author} {\bibinfo {author} {\bibfnamefont {J.}~\bibnamefont {von
  Delft}}\ and\ \bibinfo {author} {\bibfnamefont {D.}~\bibnamefont {Ralph}},\
  }\href {https://doi.org/10.1016/s0370-1573(00)00099-5} {\bibfield  {journal}
  {\bibinfo  {journal} {Physics Reports}\ }\textbf {\bibinfo {volume} {345}},\
  \bibinfo {pages} {61} (\bibinfo {year} {2001})}\BibitemShut {NoStop}%
\bibitem [{Note1()}]{Note1}%
  \BibitemOpen
  \bibinfo {note} {All perturbation calculations as well as an extended
  collection of exact numerical results is available in the form of Wolfram
  Mathematica notebooks in the Supplemental material.}\BibitemShut {Stop}%
\bibitem [{\citenamefont {Takegahara}\ \emph {et~al.}(1992)\citenamefont
  {Takegahara}, \citenamefont {Shimizu},\ and\ \citenamefont
  {Sakai}}]{takegahara1992}%
  \BibitemOpen
  \bibfield  {author} {\bibinfo {author} {\bibfnamefont {K.}~\bibnamefont
  {Takegahara}}, \bibinfo {author} {\bibfnamefont {Y.}~\bibnamefont
  {Shimizu}},\ and\ \bibinfo {author} {\bibfnamefont {O.}~\bibnamefont
  {Sakai}},\ }\href {https://doi.org/10.1143/jpsj.61.3443} {\bibfield
  {journal} {\bibinfo  {journal} {Journal of the Physical Society of Japan}\
  }\textbf {\bibinfo {volume} {61}},\ \bibinfo {pages} {3443} (\bibinfo {year}
  {1992})}\BibitemShut {NoStop}%
\bibitem [{\citenamefont {Chen}\ and\ \citenamefont {Jayaprakash}(1998)}]{CJ}%
  \BibitemOpen
  \bibfield  {author} {\bibinfo {author} {\bibfnamefont {K.}~\bibnamefont
  {Chen}}\ and\ \bibinfo {author} {\bibfnamefont {C.}~\bibnamefont
  {Jayaprakash}},\ }\href {https://doi.org/10.1103/physrevb.57.5225} {\bibfield
   {journal} {\bibinfo  {journal} {Physical Review B}\ }\textbf {\bibinfo
  {volume} {57}},\ \bibinfo {pages} {5225} (\bibinfo {year}
  {1998})}\BibitemShut {NoStop}%
\bibitem [{\citenamefont {Galpin}\ and\ \citenamefont
  {Logan}(2008{\natexlab{a}})}]{galpin2008}%
  \BibitemOpen
  \bibfield  {author} {\bibinfo {author} {\bibfnamefont {M.~R.}\ \bibnamefont
  {Galpin}}\ and\ \bibinfo {author} {\bibfnamefont {D.~E.}\ \bibnamefont
  {Logan}},\ }\bibfield  {journal} {\bibinfo  {journal} {Physical Review B}\
  }\textbf {\bibinfo {volume} {77}},\ \href
  {https://doi.org/10.1103/physrevb.77.195108} {10.1103/physrevb.77.195108}
  (\bibinfo {year} {2008}{\natexlab{a}})\BibitemShut {NoStop}%
\bibitem [{\citenamefont {Galpin}\ and\ \citenamefont
  {Logan}(2008{\natexlab{b}})}]{galpin2008epj}%
  \BibitemOpen
  \bibfield  {author} {\bibinfo {author} {\bibfnamefont {M.~R.}\ \bibnamefont
  {Galpin}}\ and\ \bibinfo {author} {\bibfnamefont {D.~E.}\ \bibnamefont
  {Logan}},\ }\href {https://doi.org/10.1140/epjb/e2008-00138-5} {\bibfield
  {journal} {\bibinfo  {journal} {The European Physical Journal B}\ }\textbf
  {\bibinfo {volume} {62}},\ \bibinfo {pages} {129} (\bibinfo {year}
  {2008}{\natexlab{b}})}\BibitemShut {NoStop}%
\bibitem [{\citenamefont {Moca}\ and\ \citenamefont {Roman}(2010)}]{Moca2010}%
  \BibitemOpen
  \bibfield  {author} {\bibinfo {author} {\bibfnamefont {C.~P.}\ \bibnamefont
  {Moca}}\ and\ \bibinfo {author} {\bibfnamefont {A.}~\bibnamefont {Roman}},\
  }\href {https://doi.org/10.1103/physrevb.81.235106} {\bibfield  {journal}
  {\bibinfo  {journal} {Physical Review B}\ }\textbf {\bibinfo {volume} {81}},\
  \bibinfo {pages} {235106} (\bibinfo {year} {2010})}\BibitemShut {NoStop}%
\bibitem [{\citenamefont {Zalom}\ \emph {et~al.}(2021)\citenamefont {Zalom},
  \citenamefont {Pokorn{\'{y}}},\ and\ \citenamefont
  {Novotn{\'{y}}}}]{Zalom2021}%
  \BibitemOpen
  \bibfield  {author} {\bibinfo {author} {\bibfnamefont {P.}~\bibnamefont
  {Zalom}}, \bibinfo {author} {\bibfnamefont {V.}~\bibnamefont
  {Pokorn{\'{y}}}},\ and\ \bibinfo {author} {\bibfnamefont {T.}~\bibnamefont
  {Novotn{\'{y}}}},\ }\href {https://doi.org/10.1103/physrevb.103.035419}
  {\bibfield  {journal} {\bibinfo  {journal} {Physical Review B}\ }\textbf
  {\bibinfo {volume} {103}},\ \bibinfo {pages} {035419} (\bibinfo {year}
  {2021})}\BibitemShut {NoStop}%
\bibitem [{\citenamefont {Zalom}\ and\ \citenamefont
  {{\v{Z}}onda}(2022)}]{Zalom2022}%
  \BibitemOpen
  \bibfield  {author} {\bibinfo {author} {\bibfnamefont {P.}~\bibnamefont
  {Zalom}}\ and\ \bibinfo {author} {\bibfnamefont {M.}~\bibnamefont
  {{\v{Z}}onda}},\ }\href {https://doi.org/10.1103/physrevb.105.205412}
  {\bibfield  {journal} {\bibinfo  {journal} {Physical Review B}\ }\textbf
  {\bibinfo {volume} {105}},\ \bibinfo {pages} {205412} (\bibinfo {year}
  {2022})}\BibitemShut {NoStop}%
\bibitem [{Note2()}]{Note2}%
  \BibitemOpen
  \bibinfo {note} {Eq.~(8) in Ref.~\protect \rev@citealpnum {CJ}, involving
  operators $f$ and $g$.}\BibitemShut {Stop}%
\bibitem [{\citenamefont {L\"{o}wdin}(1955)}]{Lowdin1955}%
  \BibitemOpen
  \bibfield  {author} {\bibinfo {author} {\bibfnamefont {P.-O.}\ \bibnamefont
  {L\"{o}wdin}},\ }\href {https://doi.org/10.1103/physrev.97.1474} {\bibfield
  {journal} {\bibinfo  {journal} {Physical Review}\ }\textbf {\bibinfo {volume}
  {97}},\ \bibinfo {pages} {1474} (\bibinfo {year} {1955})}\BibitemShut
  {NoStop}%
\bibitem [{\citenamefont {Lu}\ \emph {et~al.}(2014)\citenamefont {Lu},
  \citenamefont {H\"{o}ppner}, \citenamefont {Gunnarsson},\ and\ \citenamefont
  {Haverkort}}]{Lu2014}%
  \BibitemOpen
  \bibfield  {author} {\bibinfo {author} {\bibfnamefont {Y.}~\bibnamefont
  {Lu}}, \bibinfo {author} {\bibfnamefont {M.}~\bibnamefont {H\"{o}ppner}},
  \bibinfo {author} {\bibfnamefont {O.}~\bibnamefont {Gunnarsson}},\ and\
  \bibinfo {author} {\bibfnamefont {M.~W.}\ \bibnamefont {Haverkort}},\ }\href
  {https://doi.org/10.1103/physrevb.90.085102} {\bibfield  {journal} {\bibinfo
  {journal} {Physical Review B}\ }\textbf {\bibinfo {volume} {90}},\ \bibinfo
  {pages} {085102} (\bibinfo {year} {2014})}\BibitemShut {NoStop}%
\bibitem [{\citenamefont {He}\ and\ \citenamefont {Lu}(2014)}]{He2014}%
  \BibitemOpen
  \bibfield  {author} {\bibinfo {author} {\bibfnamefont {R.-Q.}\ \bibnamefont
  {He}}\ and\ \bibinfo {author} {\bibfnamefont {Z.-Y.}\ \bibnamefont {Lu}},\
  }\href {https://doi.org/10.1103/physrevb.89.085108} {\bibfield  {journal}
  {\bibinfo  {journal} {Physical Review B}\ }\textbf {\bibinfo {volume} {89}},\
  \bibinfo {pages} {085108} (\bibinfo {year} {2014})}\BibitemShut {NoStop}%
\bibitem [{\citenamefont {Bi}\ \emph {et~al.}(2019)\citenamefont {Bi},
  \citenamefont {Huang},\ and\ \citenamefont {Tong}}]{Bi2019}%
  \BibitemOpen
  \bibfield  {author} {\bibinfo {author} {\bibfnamefont {S.}~\bibnamefont
  {Bi}}, \bibinfo {author} {\bibfnamefont {L.}~\bibnamefont {Huang}},\ and\
  \bibinfo {author} {\bibfnamefont {N.-H.}\ \bibnamefont {Tong}},\ }\href
  {https://doi.org/10.1016/j.cpc.2018.09.002} {\bibfield  {journal} {\bibinfo
  {journal} {Computer Physics Communications}\ }\textbf {\bibinfo {volume}
  {235}},\ \bibinfo {pages} {196} (\bibinfo {year} {2019})}\BibitemShut
  {NoStop}%
\bibitem [{\citenamefont {Debertolis}\ \emph {et~al.}(2021)\citenamefont
  {Debertolis}, \citenamefont {Florens},\ and\ \citenamefont
  {Snyman}}]{debertolis2021}%
  \BibitemOpen
  \bibfield  {author} {\bibinfo {author} {\bibfnamefont {M.}~\bibnamefont
  {Debertolis}}, \bibinfo {author} {\bibfnamefont {S.}~\bibnamefont
  {Florens}},\ and\ \bibinfo {author} {\bibfnamefont {I.}~\bibnamefont
  {Snyman}},\ }\href {https://doi.org/10.1103/physrevb.103.235166} {\bibfield
  {journal} {\bibinfo  {journal} {Physical Review B}\ }\textbf {\bibinfo
  {volume} {103}},\ \bibinfo {pages} {235166} (\bibinfo {year}
  {2021})}\BibitemShut {NoStop}%
\bibitem [{\citenamefont {Schmidt}\ \emph {et~al.}(2019)\citenamefont
  {Schmidt}, \citenamefont {Benavides-Riveros},\ and\ \citenamefont
  {Marques}}]{Schmidt2019}%
  \BibitemOpen
  \bibfield  {author} {\bibinfo {author} {\bibfnamefont {J.}~\bibnamefont
  {Schmidt}}, \bibinfo {author} {\bibfnamefont {C.~L.}\ \bibnamefont
  {Benavides-Riveros}},\ and\ \bibinfo {author} {\bibfnamefont {M.~A.~L.}\
  \bibnamefont {Marques}},\ }\href {https://doi.org/10.1103/physrevb.99.224502}
  {\bibfield  {journal} {\bibinfo  {journal} {Physical Review B}\ }\textbf
  {\bibinfo {volume} {99}},\ \bibinfo {pages} {224502} (\bibinfo {year}
  {2019})}\BibitemShut {NoStop}%
\end{thebibliography}%

\end{document}